\documentclass[]{jpsj2}
\def\lsim{\buildrel {\textstyle <}\over {_\sim}}

\title{Chirality scenario of the spin-glass ordering
}
\author{Hikaru Kawamura}
\inst{Department of Earth and Space Science, Faculty of Science,
Osaka University, Toyonaka 560-0043,
Japan}
\recdate{\today}
\abst{%
Detailed account is given of the chirality scenario of experimental spin-glass transitions. In this scenario, the spin glass order of weakly anisotropic Heisenberg-like spin-glass magnets including canonical spin glasses are essentially chirality driven. Recent numerical and experimental results are discussed in conjunction with this scenario.
}

\kword{%
Chirality, chiral order, spin glass, Monte Carlo simulation
}

\begin{document}
\maketitle

\section{Introduction}

 Spin glasses are the type of random magnets in which both ferromagnetic and antiferromagnetic interactions coexist and compete, thereby giving rise to the effects of frustration and quenched randomness. In as early as 1972, certain dilute metallic alloys such as AuFe and CuMn were found to exhibit a sharp cusp-like anomaly in the susceptibility indicative of a thermodynamic transition \cite{Mydosh}. This class of randomly frustrated magnets is now called ``spin glasses''. Spin glasses have been regarded as a typical example of complex systems, and their ordering properties  have been studied quite extensively, either experimentally, analytically or numerically \cite{review}. 

 Experimentally, we now have fairly convincing evidence that spin-glass magnets exhibit an equilibrium  phase transition at a finite temperature into the glassy ordered state. This has been particularly well established in canonical spin glasses which represent a class of dilute transition-metal alloys soluted in the noble-metal host. Next question is then what is the true nature of the experimentally observed spin-glass transition and that of the low-temperature spin-glass phase. The true nature of the spin-glass ordering, however, still remains elusive and has been hotly debated \cite{review}. 

 In theoretical studies of the ordering of spin glasses, a statistical model called the Edwards-Anderson (EA) model has been widely used \cite{EA}. In this model, spins are located on a certain regular lattice, say, on a three-dimensional (3D) simple cubic lattice, and are assumed to interact via the random exchange interaction taking either ferromagnetic or antiferromagnetic sign. Note that the positional disorder of magnetic moments, which is the main source of quenched randomness in canonical spin glass, is completely neglected and is replaced by the randomness in the sign and the magnitude of the exchange interaction working on a regular lattice. 

 Spin symmetry is also an important element in the EA model. Because of its simplicity, a simple Ising model has widely been used as a ``realistic'' spin-glass model. One should bear in mind, however, that the magnetic interactions in  many spin-glass materials are nearly isotropic, being well described by an isotropic Heisenberg model, in the sense that the magnetic anisotropy is considerably weaker than the exchange interaction. In canonical spin glasses, the main exchange interaction is borne by the long-range RKKY interaction which is an isotropic interaction in spin space, while the source of the anisotropy originates from the Dzyaloshinskii-Moriya (DM) interaction or the (pseudo) dipolar interaction which is known to be weaker than the former by one or two orders of magnitude. Thus, in order to understand the properties of most of real spin-glass materials it is very important to elucidate the ordering properties of the Heisenberg EA model in 3D.

 The ordering properties of the 3D Heisenberg EA model have long been studied \cite{Banavar,McMillan,OYS,Matsubara91,Yoshino93,Kawamura92,Kawamura95,Kawamura96,Kawamura98,HukuKawa00,Matsubara00,Endoh01,Matsubara01,KawaIma,Matsumoto,Nakamura02,LeeYoung03,BerthierYoung04,ImaKawa,Picco05,HukuKawa05,Campos06,LeeYoung07,Campbell07,Kawamura07,VietKawamura,Fernandez}, and are hotly debated even now. Earlier numerical simulations on the 3D EA model suggested in common that the model exhibited only a zero-temperature transition \cite{Banavar,McMillan,OYS,Matsubara91,Yoshino93}, in apparent contrast to experiments. Common attitude in the community at that time was to invoke the weak magnetic anisotropy inherent to real materials to explain this apparent discrepancy with experiments, assuming that the weak anisotropy caused a rapid crossover from the $T_g=0$ Heisenberg behavior to the $T_g>0$ Ising behavior \cite{Bray}. Remember that the 3D Ising EA model has been known to exhibit a finite-temperature spin-glass transition \cite{Ogielski85,BhattYoung85,Ballesteros00,Katzgraber06,Jorg06,CampbellHukushima,Hasenbusch08}. In fact, however, the situation was not quite satisfactory, since the experimental exponent values measured for canonical spin glasses are actually far from the Ising spin-glass values, and no clear sign of Heisenberg-to-Ising crossover has been observed in experiments. 

 In 1992, the present author suggested that the 3D Heisenberg EA model might exhibit a finite-temperature transition {\it in its chiral sector\/} \cite{Kawamura92}. Chirality is a multispin variable representing the sense or the handedness of the noncollinear or noncoplanar structures induced by frustration, {\it i.e.\/}, whether the frustration-induced noncollinear or noncoplanar spin structure is right- or left-handed. It has subsequently been suggested that, in the ordering of the 3D Heisenberg spin glass, the chirality was ``decoupled'' from the spin, the chiral-glass order taking place at a temperature higher than the spin-glass order, $T_{CG} > T_{SG}$ \cite{Kawamura98,HukuKawa00,HukuKawa05,Kawamura07,VietKawamura}. Based on such a spin-chirality decoupling picture of the 3D isotropic Heisenberg spin glass, a chirality scenario of experimental spin-glass transition was proposed by the author \cite{Kawamura92,Kawamura07}: According to this scenario, the chirality is a hidden order parameter of real spin-glass transition. Real spin-glass transition of weakly anisotropic spin-glass magnets is then a ``disguised'' chiral-glass transition, where the chirality is mixed into the spin sector via a weak random magnetic anisotropy. 

 The chirality scenario is capable of explaining several long-standing puzzles concerning the experimental spin-glass transition in a natural way, {\it e.g.\/}, the origin of the non-Ising critical exponents experimentally observed in canonical spin glasses, and remains to be an attractive hypothesis in consistently explaining various experimental observations for canonical spin glasses. In recent numerical studies of the 3D Heisenberg EA model, although consensus now seems to appear that the 3D Heisenberg spin glass indeed exhibits a finite-temperature transition of some sort \cite{HukuKawa05,Kawamura07,Campos06,LeeYoung07,VietKawamura,Fernandez}, contrary to earlier belief in the community, the nature of the transition, especially whether the model really exhibits the spin-chirality decoupling, is still under hot debate, and the validity of the chirality scenario has been contested \cite{Matsubara00,Endoh01,Matsubara01,Nakamura02,LeeYoung03,BerthierYoung04,Picco05,Campos06,LeeYoung07,Fernandez}. 

 Under such circumstances, I wish to review in this article the present status of the chirality scenario of the spin-glass ordering in some detail, and try to further examine its consequences in terms of several recent numerical simulations and experiments.

\section{Spin glass ordering: What is at issue ?}

 In this section, I wish to argue what is at issue in the ordering of typical spin-glass magnets, canonical spin glasses in particular, and highlight some of important open questions which have remained unsolved. I wish to emphasize the three points below, either of which has experimental relevance. In fact, none of them is new, being widely recognized in the literature.

\subsection{The problem of critical properties}

 Typical spin-glass magnets exhibit an equilibrium spin-glass transition  at a  finite temperature into the spin-glass ordered phase where spins are ordered randomly without any spatial periodicity. In  many of real spin-glass materials including canonical spin glasses, the magnetic interaction is nearly isotropic, being well described by an isotropic Heisenberg model in the sense that the magnetic anisotropy is considerably weaker than the exchange interaction.  Usually, one expects for such a weakly anisotropic system that the system exhibits an anisotropy-induced crossover from the isotropic Heisenberg behavior to the anisotropic Ising behavior.

 Concerning the spin-glass critical exponents,  at least for typical canonical spin-glass materials, consistent experimental estimates are now available thanks to careful experimental measurements \cite{Fert,Simpson,Coutenary86,Bouchiat,Levy,Coles,Taniguchi88,Campbell09}. Some of these values are tabulated in the Table of the companion article by Campbell and Petit \cite{Campbell09}.  As discussed there, the exponents determined by various authors for canonical spin glasses indeed come close to each other, yielding the values $\beta \simeq 1$, $\gamma \simeq 2.2-2.3$, $\nu \simeq 1.3-1.4$ and $\eta \simeq 0.4-0.5$. As an example, we show in Fig.1 the critical behavior of the nonlinear susceptibility (corresponding to the spin-glass susceptibility) of canonical spin glass AgMn reported by Lev\'y {\it et al\/} \cite{Levy}. From the log-log plot analysis of the nonlinear susceptibility versus the reduced temperature, the associated ordering susceptibility exponent was estimated as $\gamma=2.3\pm 0.2$. 

\begin{figure}[ht]
\includegraphics[scale =0.8]{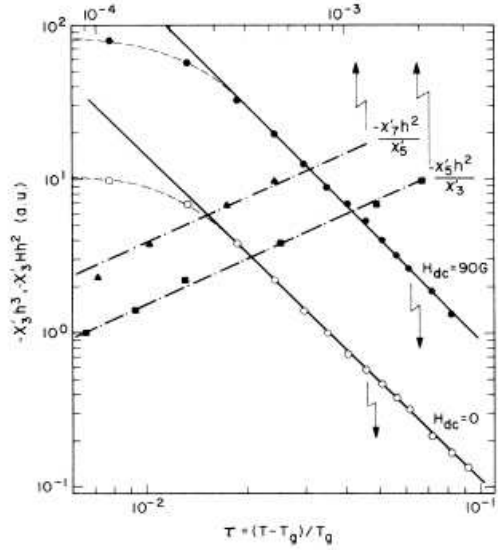}
\caption{
The temperature dependence of the nonlinear magnetic susceptibility $-\chi'_3$ of canonical spin glass AgMn. The slope gives the exponent $-\gamma$. The ratio of the higher-order susceptibilities are also given, whose slope should give the exponent $1+\beta/\gamma$. [From L.P. Lev\'y and A.T. Ogielsky, Phys. Rev. Letters {\textbf 57} (1986), 3288.]
}
\end{figure}

 We note that the corresponding exponent values of the 3D Ising EA model determined by recent extensive numerical simulations come around $\gamma \simeq 6.0\sim 6.5$, $\nu \simeq 2.5\sim 2.7$ and $\eta \simeq -0.38\sim -0.40$ \cite{CampbellHukushima,Hasenbusch08}. On comparison, one can immediately see that there is a big discrepancy between the two sets of exponents, {\it i.e.\/}, the experimental values of canonical spin glasses versus the numerical values of the 3D Ising EA model: the experimental $\nu $ and $\gamma$ are about half the numerical values and the sign of $\eta $ is reversed. Indeed, the experimental exponent values for the anisotropic Ising-like spin-glass magnet FeMnTiO$_3$ turns out to be roughly consistent with these numerical values of the 3D Ising spin glass \cite{Gunnarsson91}. Thus,  the experimental exponent values of canonical spin-glasses are clearly at variance with the 3D Ising spin-glass values.


 Another theoretical reference in studying the spin-glass critical properties should be the isotropic 3D Heisenberg EA model. Although earlier numerical results pointed toward a $T=0$ transition in apparent contrast to experiments, recent numerical simulations point to a finite-temperature spin-glass transition. Unfortunately, the exponent values associated with this $T>0$ spin-glass transition seem poorly established. The exponent $\eta$ was estimated in Ref.\cite{VietKawamura} as $\eta \sim -0.30$, while the exponent $\nu$ was estimated in Ref.\cite{Fernandez} as $\nu\sim 1.5$. The exponent $\eta$ differs significantly from the experimental value of canonical spin glasses. Hence, the experimentally observed spin-glass exponents are incompatible with either the 3D Ising nor the 3D Heisenberg spin-glass exponents. The origin of the experimentally observed spin-glass exponents has remained to be a mystery.

 One may suspect that the observed large discrepancy from the theoretical values might be caused by the long-range nature of the RKKY interaction inherent to canonical spin glasses. We note, however, that a similar discrepancy is also observed in an {\it insulating\/} Heisenberg-like spin glass with the short-range interaction, {\it e.g.\/} in thiospinel CdCr$_2$InS$_4$, where the respective exponents were obtained as $\gamma \simeq 2.3$, $\nu \simeq 1.3$ and $\eta \simeq 0.17$ \cite{Vincent}. The observed exponent values are rather close to those of metallic canonical spin glasses, being considerably different from the 3D Ising values. This observation indicates that a large deviation of the exponents from the 3D Ising values is not due to the long-range nature of the RKKY interaction, but is more or less a general attribute of the Heisenberg-like spin-glass magnets.


\subsection{The problem of phase diagram}

 Next, we proceed to the problem of the magnetic phase diagram of spin glasses.
 Unlike in the zero-field case, it remains not completely clear experimentally whether the  spin-glass ``transition'' {\it under fields\/} really exists as a true thermodynamic transition. Recent indication is that in Heisenberg-like magnets like canonical spin glasses there is a  thermodynamic spin-glass transition even in fields \cite{Campbell09,Campbell99,Campbell02}, whereas in Ising magnets like FeMnTiO$_3$ there seems to be no thermodynamic spin-glass transition in fields \cite{Nordblad95}. 

\begin{figure}[t]
\includegraphics[scale =0.3]{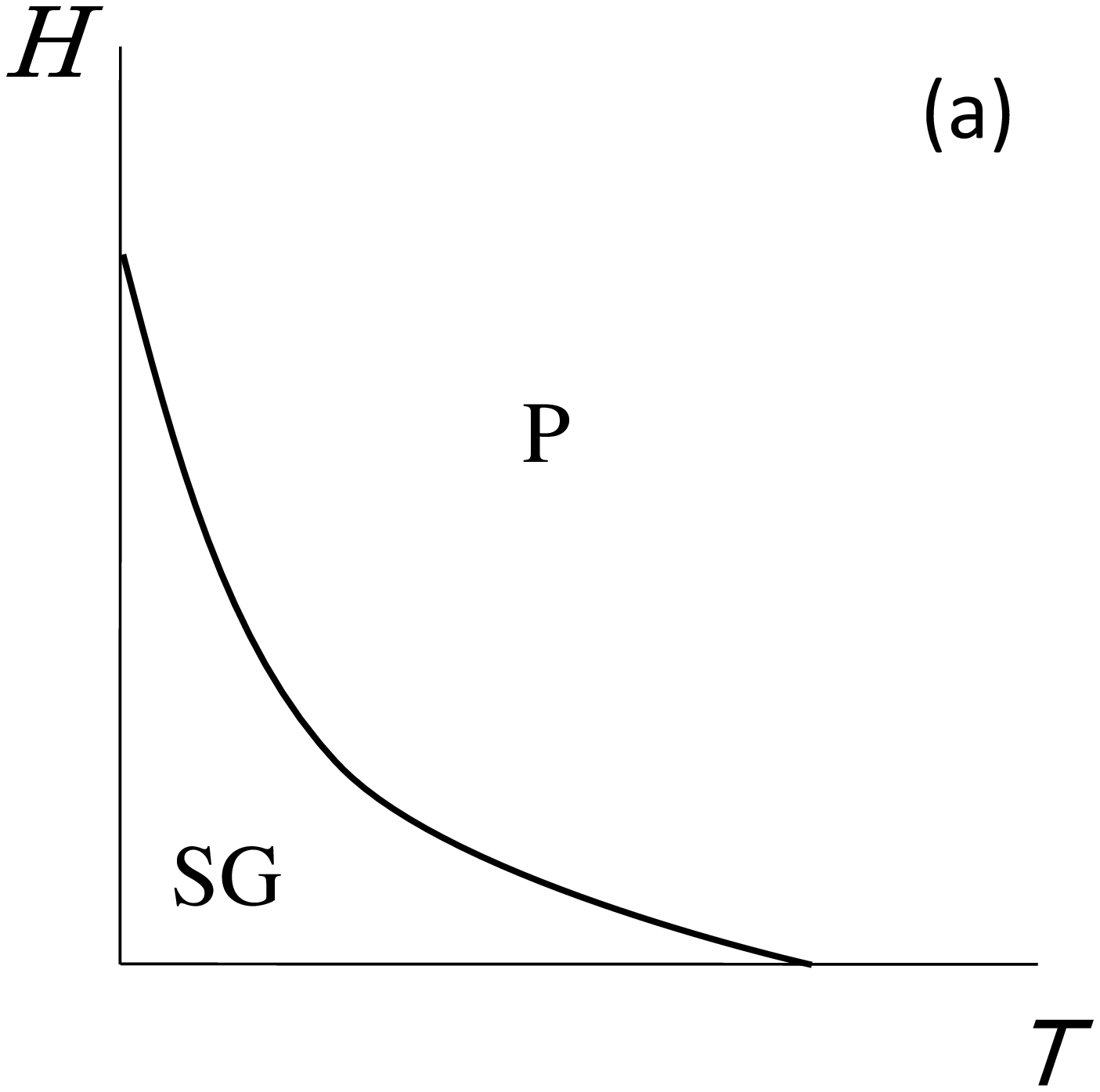}
\includegraphics[scale =0.3]{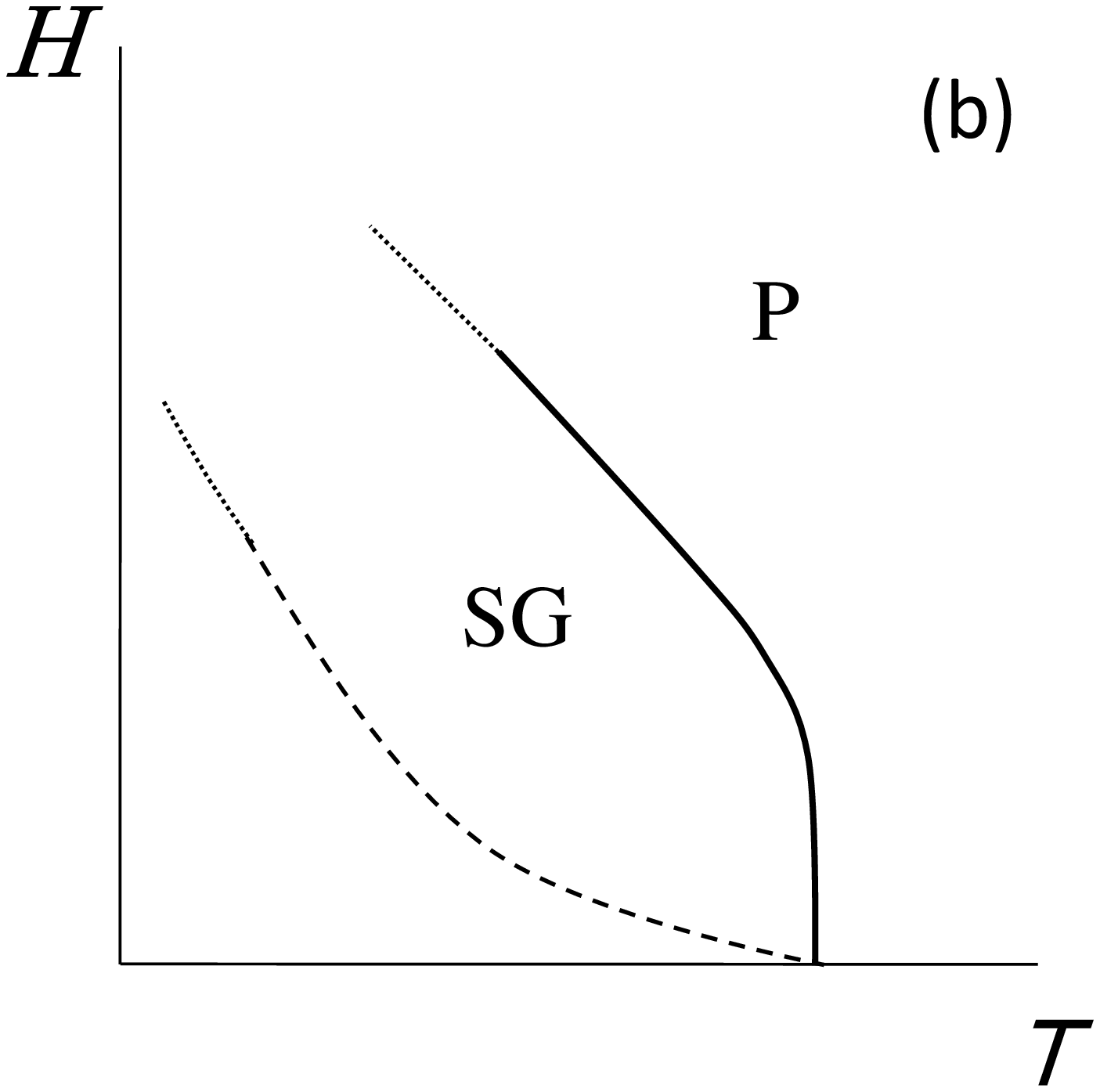}
\includegraphics[scale =0.3]{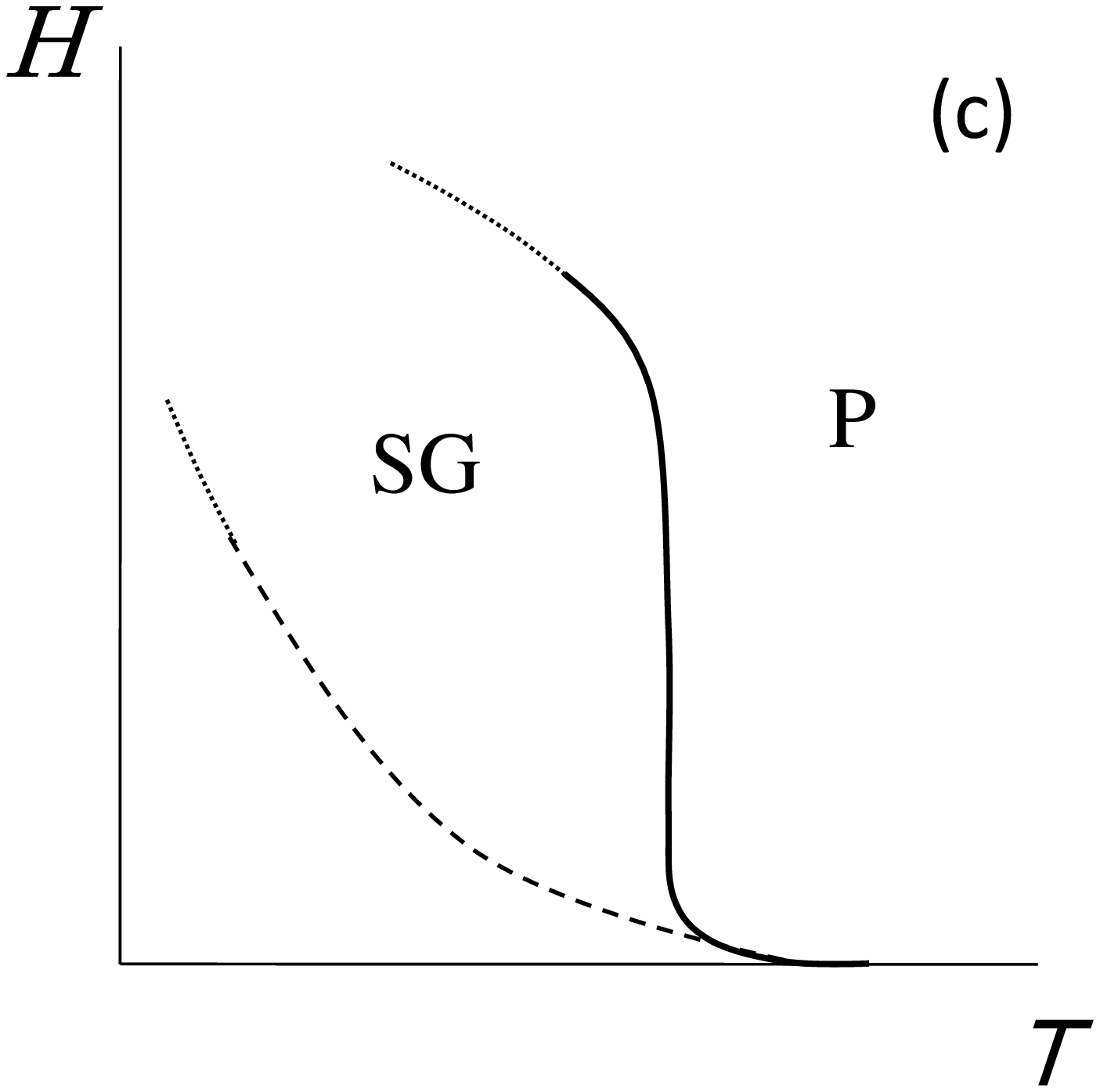}
\caption{
Mean-field phase diagrams of spin glasses in the temperature ($T)$ - magnetic field ($H$) plane for the cases of the Ising spin glass (a), the isotropic Heisenberg spin glass (b), and the weakly anisotropic Heisenberg spin glass (c). Solid line represents a true transition line, while broken line represents a crossover line. `P' stands for the paramagnetic phase and `SG' stands for the spin glass phase.
}
\end{figure}

 Meanwhile, it has widely been recognized that many of the features of the experimental phase diagram often resembles that of the mean-field spin-glass model, {\it i.e.\/}, Sherrington-Kirkpatrick (SK) model \cite{SK}. In Fig.2, we show typical mean-field phase diagrams for each case of the Ising spin glass (a), the isotropic Heisenberg spin glass (b), and the Heisenberg-like spin glass with weak random anisotropy (c). In the Ising case (a), the transition line under fields is the so-called de Almeida-Thouless (AT) line associated with a spontaneous replica-symmetry breaking (RSB),  which behaves as $H \approx |T_g(H)-T_g(0)|^{3/2}$ as $H\rightarrow 0$ \cite{AT}. In the isotropic case (b), by contrast, the transition line under fields is the so-called the Gabay-Toulouse (GT) line, associated with the onset of the transverse spin order, which behaves as $H \approx |T_g(H)-T_g(0)|^{1/2}$ as $H\rightarrow 0$ \cite{GT}. The AT-line associated with the onset of the longitudinal spin order also appears as a crossover line (not a true transition line) at a lower temperature. In the weakly anisotropic case (c), which has the most direct relevance to real Heisenberg-like spin-glass magnets, the AT-line appears at lower fields as a true transition line behaving as $H \approx |T_g(H)-T_g(0)|^{3/2}$, while at higher fields it changes over to the GT line behaving as $H \approx |T_g(H)-T_g(0)|^{1/2}$ \cite{Kotliar}. Even at higher fields, the AT line remains at a lower temperature as a crossover line, which is a continuation of the AT line at lower fields.

 The experimental phase diagram of the weakly anisotropic Heisenberg spin glasses like AuFe, CuMn and AgMn often looks quite similar to the mean-field phase diagram of the weakly anisotropic Heisenberg spin glass, Fig.2(c) \cite{Campbell09,Campbell99,Campbell02,Chamberlin,Courtenary84,Kennig}.  Such coincidence with the mean-field phase diagram including the exponents describing the phase boundary, however, is a bit surprising if one notices the fact that the mean-field theory usually gives a poor result on the exponent. Indeed, concerning the zero-field susceptibility exponent $\gamma$, a mean-field theory yields $\gamma=1$, while the corresponding experimental value  for canonical spin glasses is known to be $\gamma\simeq 2.2$. The experimental value and the mean-field value are not necessarily close. Thus, it remains puzzling why the exponents describing the in-field phase boundary, {\it i.e.\/}, 3/2 for the AT-line and 1/2 for the GT-line, agree so well with the experimental values.



\subsection{The problem of replica-symmetry breaking}

 One of hot issues in the spin-glass research has been concerned with the nature of the spin-glass ordered state: In particular, whether the spin-glass ordered state spontaneously breaks replica symmetry or not \cite{review}. Two typical views have been common. One is a droplet picture, which claims that the spin-glass ordered state is a ``disguised ferromagnet'' without a spontaneous RSB \cite{FisherHuse}. The other is a hierarchical RSB picture inspired by the exact solution of the mean-field model, which claims that the spin-glass ordered state is intrinsically more complex accompanied with a hierarchical or full RSB where the phase space is hierarchically organized in the spin-glass ordered state \cite{Parisi}. Hot debate has continued over years concerning which view applies to the ordered state of real spin-glass magnets. 

 For more quantitative discussion, it is convenient to introduce an ``overlap'' variable $q$, which is defined by
\begin{equation}
q=\frac{1}{N}\sum_{i} S_i^{(\alpha)}S_i^{(\beta)}
\end{equation}
where $S_i^{(\alpha)}$ represents the $i$-th spin variable of the ``replica'' $\alpha$ and the summation is taken over all $N$ spins of the system. Replicas $\alpha$ and $\beta$ mean here the two independent copies of the system with the same realization of quenched randomness. One can then consider the distribution function of the overlap variable $P(q)$ 
\begin{equation}
P(q')=[<\delta(q-q')>]
\end{equation}
where $<\cdots>$ represents a thermal average and [$\cdots$] an average over the quenched disorder (configurational average). Some typical forms of $P(q)$ in the thermodynamic limit is illustrated in Fig.3.

\begin{figure}[t]
\includegraphics[scale =0.35]{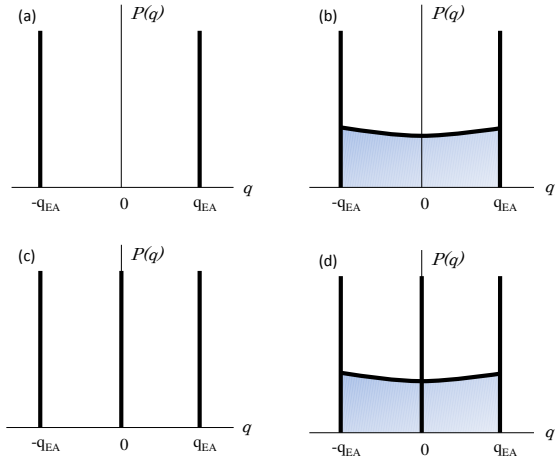}
\caption{
(Color online) Typical patterns of the overlap distribution function $P(q)$ in the thermodynamic limit. In the case (a) expected in a droplet theory, $P(q)$ consists of just two delta-function peaks at $q=\pm q_{EA}$. In the case of (b) expected in a hierarchical or full RSB picture, $P(q)$ possesses  an additional plateau part connecting the two delta-function peaks at $q=\pm q_{EA}$. In the case of (c) expected in a one-step RSB picture, $P(q)$ possesses  a central peak located at $q=0$ in addition to two delta-function peaks at $q=\pm q_{EA}$. The case (d) is a combination of (b) and (c).
}
\end{figure}

 The droplet picture claims that the overlap distribution describing the spin-glass ordered state  to be a trivial one  in the thermodynamic limit consisting of just two delta functions located at $q=q_{EA}$ and at $q=-q_{EA}$. It means that the spin-glass ordered state consists of unique pure state and its symmetry counterpart, irrespective of its apparent complexity in real-space spin pattern. 

 In the hierarchical RSB picture, by contrast, $P(q)$ exhibits a continuous plateau part spanning between the two delta-function peaks at $q=\pm q_{EA}$. It means that the phase space is divided into infinitely many pure states organized in a hierarchical manner, each of which is separated by infinitely high free-energy barrier. 
 
  Other types of RSB have also been known. One well-known example might be a one-step RSB, in which $P(q)$ possesses a central $\delta$-function peak at $q=0$ in addition to the self-overlap peaks  at $q=\pm q_{EA}$. In this case, the phase space is divided into many components, but all of them, except for itself and its symmetry partner, are completely dissimilar or orthogonal. It is realized, {\it e.g.\/}, in the ordered states of the mean-field Potts glass or of the mean-field $p$-spin model \cite{review,HukuKawaMF,Picco01}. In the past, such a one-step RSB has often been discussed in the context of the structural-glass problem rather than spin-glass problem. The combination of the full RSB and the one-step RSB is also possible in certain models.

\begin{figure}[t]
\includegraphics[scale =0.37]{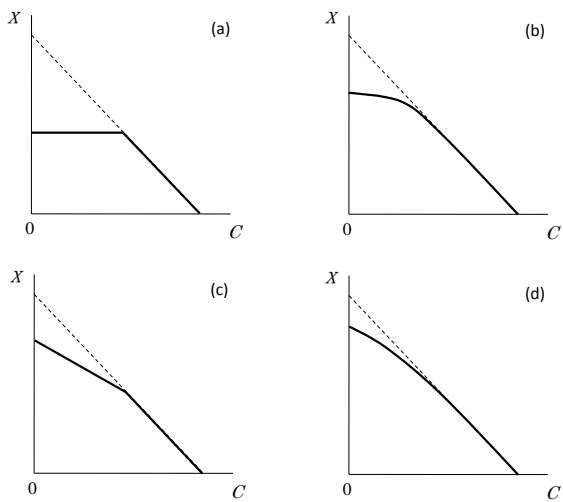}
\caption{
Typical patterns of off-equilibrium susceptibility $X$ versus correlation $C$ plot expected in the long waiting-time limit. The cases (a)-(d) correspond to the cases of (a)-(d) of Fig.3. The function $X(C)$ is related to the overlap distribution function $P(q)$ by the relation $P(q)=\frac{{\rm d}X(q)}{{\rm d}q}$.
}
\end{figure}

 In the situation that no direct experimental information about the overlap distribution function $P(q)$ has been available, the issue of whether the spin-glass ordered state accompanies RSB or not has remained largely a pure theoretical one. However, recent progress on off-equilibrium dynamical properties of spin glasses has changed this situation considerably, and has opened the door to serious experimental exploration of $P(q)$ and RSB \cite{review-off}. 

 One promising way toward this goal is to measure the so-called fluctuation-dissipation ratio. In equilibrium, there holds a well-known relation between the response and the correlation of the system known as the fluctuation-dissipation theorem (FDT). In off-equilibrium, relaxation of physical quantities of spin glasses depends on its previous history. Most typically, it depends not only on the observation time $t$ but also on the waiting time $t_w$, {\it i.e.\/}, exhibits aging. While relaxation is still stationary and FDT holds in the short-time quasi-equilibrium regime $t_0<<t<<t_w$ ($t_0$ is a microscopic time scale), it becomes non-stationary and FDT is broken in the long-time aging regime $t>>t_w$. 

 The breaking pattern of FDT is described by the so-called fluctuation-dissipation ratio (FDR) $X$, which is defined by the relation,
\begin{equation}
R(t_1,t_2)=\frac{X(t_1,t_2)}{k_BT}\frac{\partial C(t_1,t_2)}{\partial t_1},
\end{equation}
where $R(t_1,t_2)$ is a response function measured at time $t_2$ to an impulse field applied at time $t_1$, $C(t_1,t_2)$ is a two-time correlation function in zero field at times $t_1$ and $t_2=t_1+t$, and $T$ is the bath temperature. One may regard $T/X\equiv T_{{\rm eff}}$ as an effective temperature. In the case FDT holds, one has $X=1$ and $T_{{\rm eff}}=T$. Recent studies have revealed that in the limit of infinite time $t_1\rightarrow \infty$, the FDR $X$ depended on the times $t_1$ and $t_2$ only through the correlation function $C(t_1,t_2)$, {\it i.e.\/}, $X(t_1,t_2)=X(C(t_1,t_2))$ \cite{CK}, and that $X(C)$ is equivalent to the so-called $x(q)$-function, an integral of the overlap distribution function $P(q)$, $P(q)=\frac{{\rm d}x(q)}{\rm{d}q}$ \cite{CK}. If this is the case, the information about $P(q)$ or $x(q)$ is obtainable by measuring both the response or the susceptibility $\chi$ and the correlation $C$, with the time $t$ as an implicit parameter. In Fig.4, we illustrate several typical examples of the $\chi$ vs. $C$ plot expected for four typical behaviors of the RSB pattern shown in Fig.3.

\begin{figure}[t]
\includegraphics[scale =0.8]{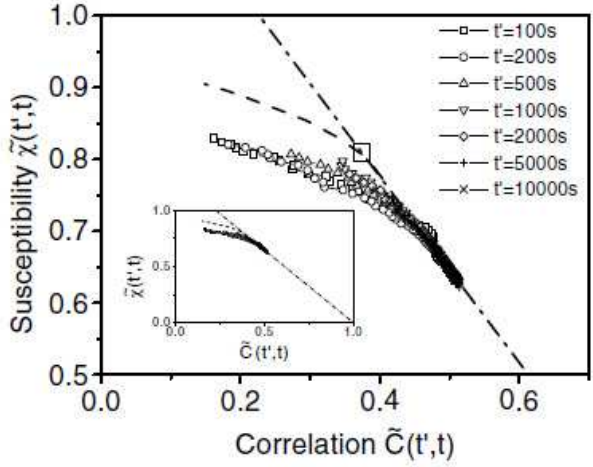}
\caption{
 Plot of off-equilibrium susceptibility $\tilde \chi$ versus correlation $\tilde C$ of an insulating Heisenberg-like spin glass, thiospinel CdCr$_{1.7}$In$_{0.3}$S$_4$. The dot-dashed line is the FDT line. The temperature is $T=0.8T_g=13.3$K, while $t'$ represents the waiting time. The dashed line in the figure represents an extrapolation to $t'\rightarrow \infty$. [From D. H\'erisson and M. Ocio {\it et al\/}, Phys. Rev. Letters {\textbf 88} (2002) 257202.]
}
\end{figure}

  H\'erisson and Ocio experimentally measured the $\chi$-$C$ relation for an insulating Heisenberg spin glass, thiospinel CdCr$_{1.7}$In$_{0.3}$S$_4$, giving experimental information about the FDR of spin glass \cite{HO}. The $\chi$ vs. $C$ plot reported in Ref.\cite{HO} is reproduced in Fig.5.  On comparison with Fig.4, the experimental result seems incompatible with the one of Fig.4(a), indicating the occurrence of RSB in the spin-glass ordered state of thiospinel. Identification of the RSB pattern seems more subtle due to the lack of the data near $C=0$, but the behavior seems indicative of the pattern of Fig.4(c) or (d), {\it i.e.\/}, a certain feature of one-step-like RSB. 

 We note that other proposal was also made to experimentally detect $P(q)$. For example, Carpentier recently proposed that the overlap $q$ of mesoscopic metallic spin glasses might be measurable via the conductance fluctuations \cite{Carpentier}. In any case, the direct experimental information about the RSB pattern of the spin-glass ordered state  now seems to be within our reach.

\medskip
 In concluding this section, we wish to summarize the main issues concerning the experiments on spin glasses  which remains to be puzzling and needs some explanation. They are:

\medskip\noindent
i) Why the exponents observed in Heisenberg-like spin glasses, including both metallic and insulating, largely deviate from the 3D Ising values ? What is the origin of these exponents observed quite in common for Heisenberg-like spin-glass magnets ?

\medskip\noindent
ii) Why the mean-field phase diagram often gives such a good description of the phase diagram of real experimental Heisenberg-like spin glasses ? 

\medskip\noindent
iii) Does RSB occur in the spin-glass ordered state ? If yes, what type ?

\section{Chirality scenario of experimental spin-glass ordering}

In the present section, I wish to review and further develop chirality scenario of experimental spin-glass ordering of Heisenberg-like spin-glass magnets \cite{Kawamura92,Kawamura07}.

\subsection{Chirality}

 We begin with the definition of the local chirality variable. While two types of chirality have often been discussed in the literature, a vector chirality and a scalar chirality, the one relevant to the Heisenberg-like spin glass is the latter, {\it i.e.\/}, the scalar chirality. The three-component Heisenberg spin system ordered in a noncoplanar manner under the isotropic exchange interaction possesses a twofold $Z_2$ chiral degeneracy, according as the noncoplanar spin structure is either right- or left-handed, in addition to the $SO(3)$ spin-rotation degeneracy. The scalar chirality $\chi$ is defined by the product of three neighboring spins by 
\begin{equation}
\chi = \vec S_i\cdot  \vec S_j\times  \vec S_k.
\end{equation}

\subsection{Overview of the scenario}

 The chirality scenario consists of the two parts, {\it i.e.\/}, the ``spin-chirality decoupling'' part for a completely isotropic system and the ``spin-chirality recoupling'' part for a weakly anisotropic system. The first part, the spin-chirality decoupling, is a key ingredient of the scenario. It claims that the fully isotropic 3D Heisenberg spin glass exhibits a peculiar two-step ordering process, in which the systems exhibits, with decreasing the temperature, first the glass ordering of the chirality at a finite temperature $T=T_{CG}$ spontaneously breaking only a discrete $Z_2$ symmetry with preserving the continuous $SO(3)$ symmetry, and at a lower temperature $T=T_{SG}<T_{CG}$ exhibits the glass ordering of the spin itself fully breaking both the $Z_2$ and $SO(3)$ symmetries. The higher transition at $T=T_{CG}$ associated with the discrete $Z_2$ symmetry breaking is called the ``chiral-glass transition'', while the intermediate phase between $T_{CG}$ and $T_{SG}$ where only the $Z_2$ symmetry is broken with preserving the continuous $SO(3)$ symmetry is called the ``chiral-glass state''.  

 The second part of the chirality scenario concerns the role of weak anisotropy which inevitably exists in real spin-glass magnets to certain extent. The chirality scenario claims that the weak random magnetic anisotropy, which reduces the Hamiltonian symmetry from $Z_2\times SO(3)$ to only $Z_2$, weakly ``mixes'' the chirality to the spin sector, and the chiral-glass transition hidden in the chiral sector in fully isotropic system is `revealed' in the spin sector in weakly anisotropic system. In this scenario, the chiral-glass transition of the fully isotropic system, not the spin-glass transition of the isotropic system, dictates the spin-glass of the weakly anisotropic real spin glasses.

 Below, I will describe the scenario in a bit more detail.

\subsection{Spin-chirality decoupling of isotropic system}

 We first discuss the spin-chirality decoupling in the fully isotropic system. We stress that such a decoupling phenomenon, if any, is a long-scale phenomenon. As is evident from the definition of the local chirality, the chirality is a composite operator of the spins locally, not independent of the spin. Namely, at short length scale of order lattice spacing, the chirality is trivially coupled to the spin as $\chi\sim S^3$. The spin-chirality decoupling, if any, means that, on sufficient long length and time scales, say, beyond a certain crossover length and time scale, chiral correlations might outgrow spin correlations, {\it i.e.\/}, the chiral correlation length gets much longer than the spin correlation length, $\xi$ (chirality) $>>$ $\xi$(spin).

 The basic picture is summarized in Fig.6 in terms of the temperature dependence of the spin and the chirality correlation lengths. With decreasing the temperature, chiral correlations outgrow spin correlations at some crossover temperature $T=T_\times$, and the chirality exhibits a glass transition at a temperature $T=T_{{\rm CG}}$ into the {\it chiral-glass\/} ordered state without accompanying the standard spin-glass order. The spin-glass transition sets in at a temperature lower than the chiral-glass transition temperature, $T_{{\rm SG}}<T_{{\rm CG}}$, and the chiral-glass phase is realized between $T_{CG}$ and $T_{SG}$.

\begin{figure}[t]
\includegraphics[scale =0.5]{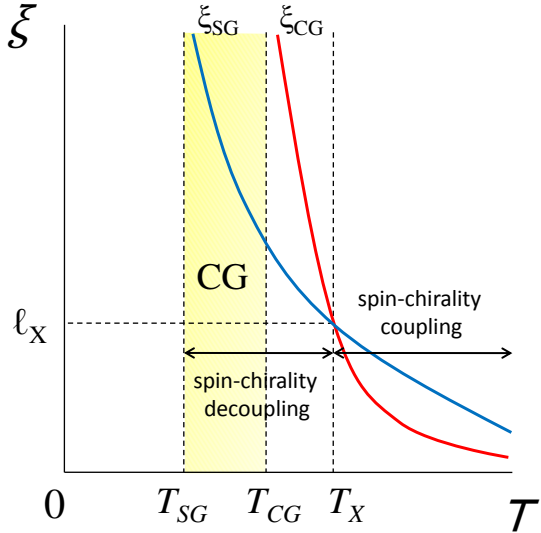}
\caption{
(Color online) The temperature dependence of the correlation length $\xi$ of the 3D Heisenberg spin glass both for the spin and for the chirality, expected from the chirality scenario. The shaded region represents the chiral-glass phase. $T_{CG}$, $T_{SG}$ and $T_\times$ denote the chiral-glass, the spin-glass and the crossover temperatures, respectively, while $l_\times$ denotes the crossover length.
}
\end{figure}

 The issue of whether the spin-chirality decoupling really occurs in the 3D Heisenberg spin glass has remained controversial for years. While several numerical results in favor of the occurrence of the spin-chirality decoupling were reported \cite{Kawamura98,HukuKawa00,KawaIma,Matsumoto,ImaKawa,HukuKawa05,VietKawamura,Fernandez}, a simultaneous spin and chirality transition without the spin-chirality decoupling was claimed in other works \cite{Matsubara00,Endoh01,Matsubara01,Nakamura02,LeeYoung03,BerthierYoung04,Picco05,Campos06,LeeYoung07,Fernandez}. The recent simulation of Ref.\cite{VietKawamura}, however, has provided a fairly strong numerical support for the occurrence of the spin-chirality decoupling. This calculation indicates that the spin-glass transition point $T_{SG}$ is located about 10\% $\sim$ 15\% below the chiral-glass transition point $T_{CG}$.

\begin{figure}[ht]
\begin{center}
\includegraphics[scale=0.9]{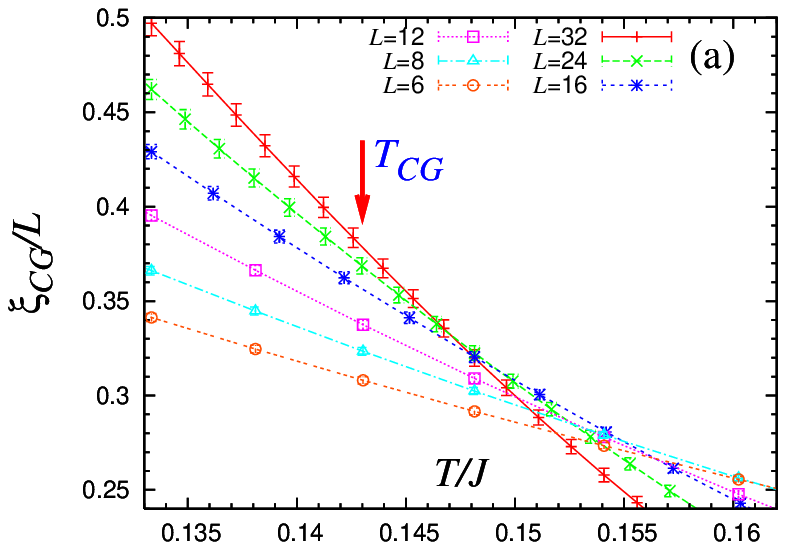}
\includegraphics[scale=0.9]{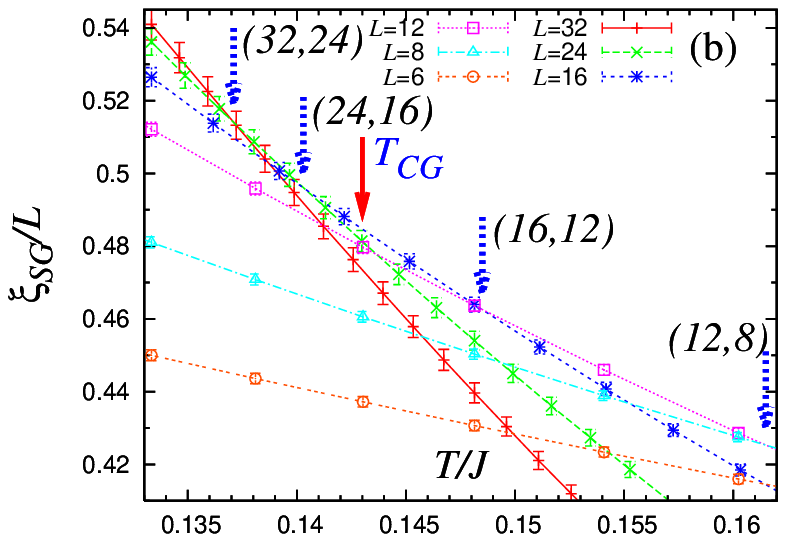}
\end{center}
\caption{
(Color online) The temperature and size dependence of the correlation-length ratio for the chirality (a), and for the spin (b). The arrow indicates the bulk chiral-glass transition point. [From D.X. Viet and H. Kawamura: Phys. Rev. B{\bf 80} (2009) 064118.]
}
\end{figure}
\begin{figure}[ht]
\begin{center}
\includegraphics[scale=0.9]{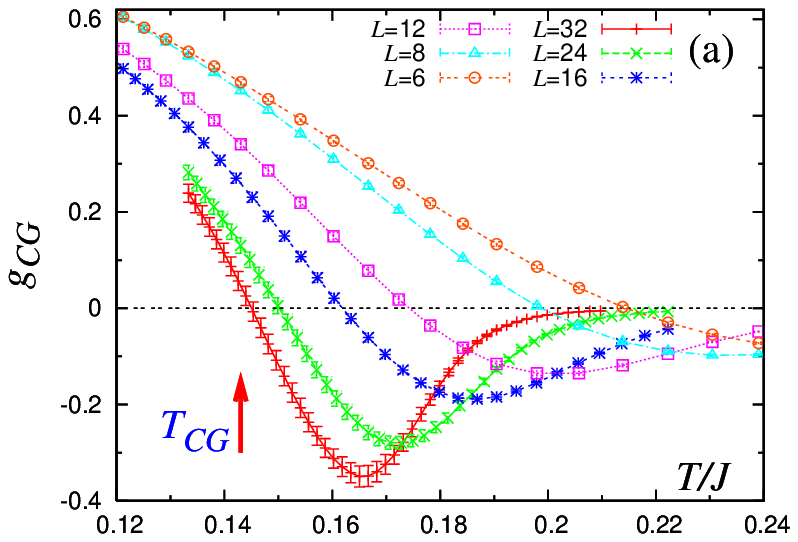}
\includegraphics[scale=0.9]{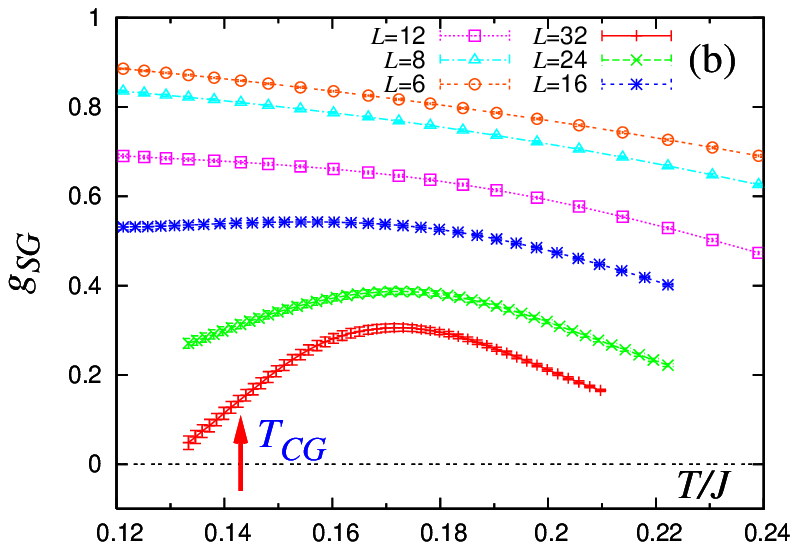}
\end{center}
\caption{
(Color online) The temperature and size dependence of the Binder ratio for the chirality (a), and for the spin (b). The arrow indicates the bulk chiral-glass transition point.  [From D.X. Viet and H. Kawamura: Phys. Rev. B{\bf 80} (2009) 064118.]
}
\end{figure}
\begin{figure}[ht]
\begin{center}
\includegraphics[scale=0.9]{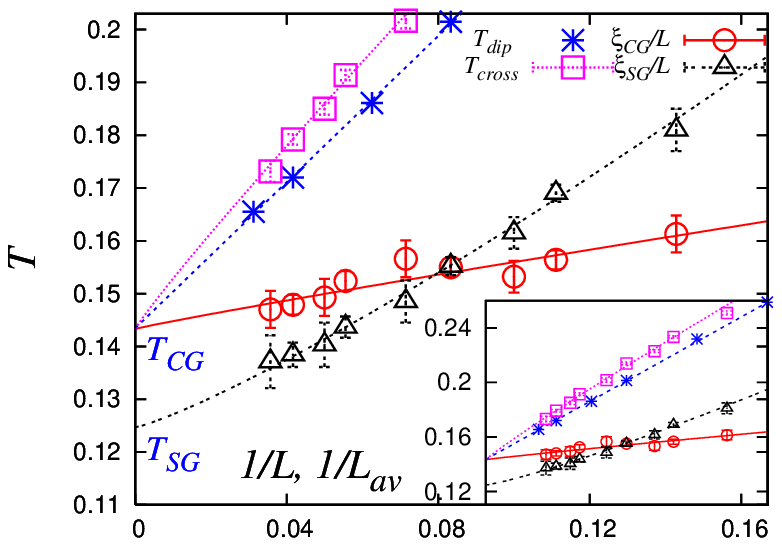}
\end{center}
\caption{
(Color online) The (inverse) size dependence of the crossing temperatures of $\xi_{CG}/L$ and  $\xi_{SG}/L$, the dip temperature $T_{dip}$ and the crossing temperature $T_{cross}$ of $g_{CG}$. The inset exhibits a wider range.  [From D.X. Viet and H. Kawamura: Phys. Rev. B{\bf 80} (2009) 064118.]
}
\end{figure}
 In Fig.7, we show the recent Monte Carlo (MC) data of the correlation-length ratios for the chirality  $\xi_{CG}/L$ (a), and for the spin  $\xi_{SG}/L$ (b), of the isotropic 3D Heisenberg EA model with random Gaussian coupling \cite{VietKawamura}. The system contains total $N=L^3$ spins with $L$ ranging from 6 to 32. The correlation-length ratios are dimensionless quantities so that the data of various $L$ should be scale-invariant and exhibits a crossing behavior at the respective spin-glass and chiral-glass transition points. For further details, refer to Ref.\cite{VietKawamura}. As can be seen from the figure, while the chiral $\xi_{CG}/L$ curves cross at temperatures which are only weakly $L$-dependent, the spin $\xi_{SG}/L$ curves cross at progressively lower temperatures as $L$ increases.  

 As an other indicator of the transition, we show in Fig.8 the Binder ratios for the chirality (a), and for the spin (b). The Binder ratios are also dimensionless, and are expected to exhibit a scale-invariant behavior at the respective chiral-glass and spin-glass transition points. As can be seen from the figure, the chiral Binder ratio $g_{CG}$ exhibits a negative dip which deepens with increasing $L$. The data of different $L$ cross on the {\it negative\/} side of $g_{CG}$ unlike the correlation-length ratio. These features indicate a finite-temperature transition in the chiral sector. 

 To estimate the bulk chiral-glass and spin glass transition temperatures quantitatively, we plot in Fig.9 the crossing temperature of $\xi_{CG}/L$ and $\xi_{SG}/L$ for pairs of successive $L$ values versus $1/L_{av}$, where $L_{av}$ is a mean of the two sizes, together with  the dip temperature $T_{dip}$ and the crossing temperature $T_{cross}$ of the chiral Binder ratio  $g_{CG}$. The data show near-linear $1/L_{av}$-dependence. The chiral crossing temperatures of  $\xi_{CG}/L$ and of $g_{CG}$ exhibits a weaker size dependence than the spin crossing temperature, and are extrapolated to $T_{CG}= 0.143\pm 0.003$ (in units of $J$). The spin crossing temperature exhibits a stronger size dependence, which is extrapolated to $T_{SG}= 0.125^{+0.006}_{-0.012}$. Hence, $T_{SG}$ is lower than $T_{CG}$ by about 10\% $\sim$ 15\%.

\begin{figure}[t]
\includegraphics[scale =1.0]{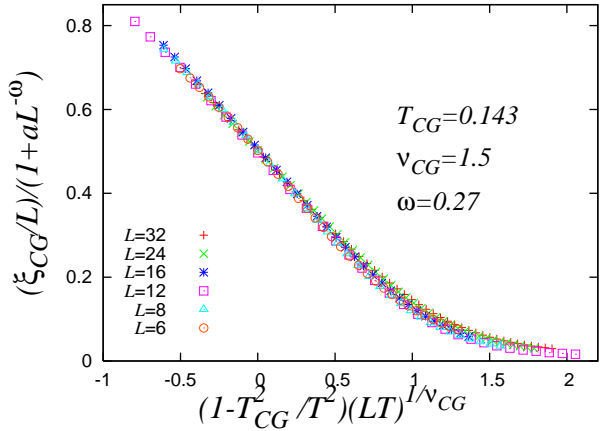}
\caption{
(Color online) Finite-size-scaling plot of the chiral-glass correlation-length ratio $\xi_{CG}/L$, where the correction-to-scaling effect is taken into account. The chiral-glass transition temperature and the leading correction-to-scaling exponents are  $T_{CG}=0.143$ and $\omega=0.27$. The best fit is obtained with $\nu_{CG}=1.5$.  [From D.X. Viet and H. Kawamura: Phys. Rev. B{\bf 80} (2009) 064118.]
}
\end{figure}

 Various chiral-glass exponents have been estimated via the standard finite-size scaling analysis. As an example, we show in Fig.10 the finite-size scaling plot of the chiral correlation-length ratio $\xi_{CG}/L$ in which the effect of the leading correction-to-scaling term has been taken into account.  The scaling is done here according to the method of Ref.\cite{CampbellHukushima}, where the scaling variable is chosen so that the data in the scaling regime match those in the high-temperature regime to get an extended scaling regime. The exponents determined in this way from various physical quantities yield $\nu_{CG}= 1.5\pm 0.2$ and $\eta_{CG}= 0.6\pm 0.2$, {\it etc.\/}, which differ significantly from the standard 3D Ising spin-glass values, $\nu \simeq 2.5\sim 2.7$ and $\eta \simeq -0.38\sim -0.40$ \cite{CampbellHukushima,Hasenbusch08}. The results indicate that the chiral-glass transition belongs to a universality class different from that of the 3D Ising spin glass. Possible long-range and/or many-body nature of the chirality-chirality interaction might be the cause of this difference.

 Other interesting issue concerns with the nature of the chiral-glass ordered state, {\it i.e.\/}, whether the chiral-glass state exhibits an RSB, and if so, what type. In this connection, it should be noticed that the form of the chiral Binder ratio $g_{CG}$ shown in Fig.9, which exhibits a prominent negative dip, is quite peculiar. In fact, this form of the Binder ratio resembles the one of the system exhibiting a one-step RSB \cite{HukuKawaMF,Picco01}. 

 In Fig.11, we show the chiral-overlap distribution $P(q_\chi)$ in the chiral-glass phase  calculated in Ref.\cite{HukuKawa05} for the 3D Heisenberg spin-glass model with the binary coupling. Here, the chiral overlap $q_\chi$ is defined by $q_\chi=(1/N)\sum q_i^{(1)}q_i^{(2)}$, where (1) and (2) indicate two copies (replicas) of the system. The calculated $P(q_\chi)$ exhibits besides symmetric side peaks located at $q_\chi =\pm q_\chi^{{\rm EA}}$ corresponding to the long-range chiral-glass order, which grow with increasing $L$,  it also exhibits a prominent central peak at $q_\chi =0$, which also grows with increasing $L$. The existence of such a pronounced central peak is a characteristic feature of the system exhibiting a one-step-like RSB, never seen in the Ising spin glass. The data strongly suggest that the chiral-glass ordered state exhibits a one-step-like RSB \cite{HukuKawa00,HukuKawa05,VietKawamura}. Recent MC also indicates that the chiral-glass ordered state is non-self-averaging \cite{HukuKawa05,VietKawamura}.

\begin{figure}[t]
\includegraphics[scale =0.55]{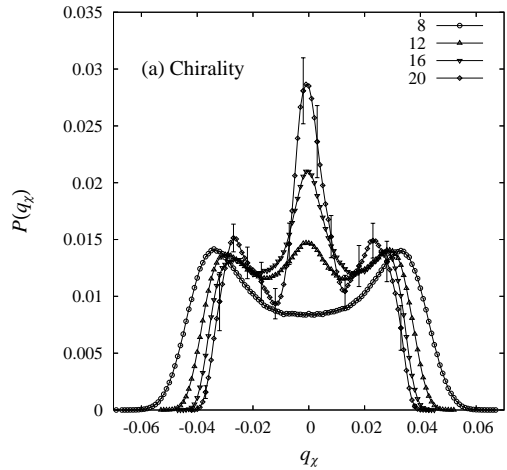}
\caption{
The overlap distribution function for the chirality of the 3D Heisenberg spin glass with the $\pm J$ binary coupling.  The temperature is $T=0.15$ below the chiral-glass transition temperature of this model $T=T_{CG}\simeq 0.19$.  [From K. Hukushima and H. Kawamura, Phys. Rev. B{\textbf 72} (2005) 144416.]
}
\end{figure}

\subsection{Spin-chirality recoupling of weakly anisotropic system}

 On assuming that the spin-chirality decoupling occurs in the 3D isotropic Heisenberg spin glass, we now wish to move to the latter part of the chirality scenario, {\it i.e.\/}, the spin-chirality recoupling due to the weak random magnetic anisotropy. As mentioned, the chirality scenario claims that the weak random anisotropy inherent to real spin-glass magnets ``recouples'' the spin to the chirality, and the chiral-glass transition of the isotropic system is revealed as the standard spin-glass transition in real weakly anisotropic Heisenberg spin glass \cite{Kawamura92,Kawamura07}. 

 Such a ``spin-chirality recoupling'' can be understood based on a simple symmetry consideration. The isotropic Heisenberg spin glass possesses both the chiral $Z_2$ symmetry and the spin-rotation $SO(3)$ symmetry, {\it i.e.\/}, $Z_2 \times SO(3)$. Due to the spin-chirality decoupling, only the chiral $Z_2$ is spontaneously broken in the isotropic system at the chiral-glass transition $T=T_{CG}$ with keeping the $SO(3)$ symmetry unbroken, which leaves the spin to be paramagnetic even below $T_{{\rm CG}}$. Suppose that the weak random anisotropy is now added to the isotropic system. It energetically breaks the $SO(3)$ symmetry with keeping the chiral $Z_2$ symmetry. (Note that the invariance under  the time reversal or the spin inversion, $S \rightarrow - S$, which flips the chirality, is kept in the presence of the random magnetic anisotropy.)  Since the chiral $Z_2$ has already been decoupled from the $SO(3)$ in the isotropic system, the $Z_2$ chiral-glass transition would persist even in the anisotropic system essentially in the same manner as that of the isotropic system. As soon as the $Z_2$ chiral-glass transition takes place, however, there is no longer any global symmetry left in the anisotropic system, which forces the spin to order below $T_{{\rm CG}}$. In other words, the random anisotropy works as an effective random field acting on the decoupled $SO(3)$ part, preserving the decoupled chiral $Z_2$ part intact. This is a spin-chirality recoupling due to the random magnetic anisotropy. 

\begin{figure}[t]
\includegraphics[scale =0.45]{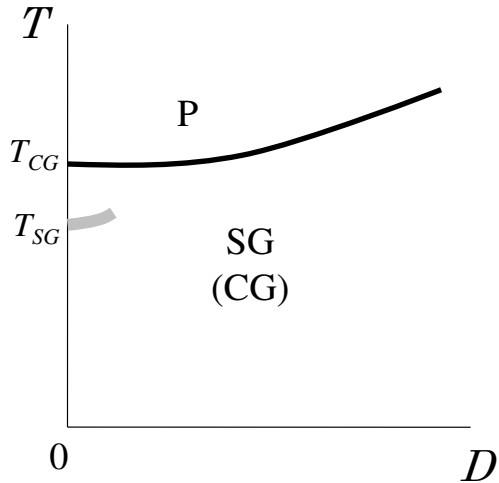}
\caption{
Schematic phase diagram of the weakly anisotropic Heisenberg spin glass in the anisotropy ($D$) versus temperature ($T$) plane. `P', `CG' and `SG'  stand for the paramagnetic, chiral-glass and spin-glass phases, respectively. The solid line represents a true transition line, while the thin shaded line represents a crossover line.
}
\end{figure}

 This situation is sketched in a schematic phase diagram in the anisotropy ($D$) versus temperature ($T$) plane of Fig.12. In the isotropic limit $D=0$, due to the spin-chirality decoupling, the chiral-glass transition occurs at a temperature higher than the spin-glass transition temperature, $T_{{\rm CG}} > T_{{\rm SG}}$. A crucial observation here is that the spin glass (simultaneously chiral-glass) transition of the anisotropic $D>0$ system is a continuation of the chiral-glass fixed point of the isotropic $D=0$ system, {\it not\/} a continuation of the $D=0$ spin-glass fixed point. Then, the spin-glass transition of real Heisenberg-like spin-glass magnets with weak random anisotropy is governed by the same $D=0$ chiral-glass fixed point along the transition line including both $D=0$ and $D>0$. In this way, the $D\rightarrow 0$ limit is not necessarily singular, not accompanying Heisenberg-to-Ising crossover in the spin-glass critical properties in the $D\rightarrow 0$ limit.

 If the $D\rightarrow 0$ limit of the chiral-glass transition is non-singular, the anisotropy dependence of the spin-glass (simultaneously chiral-glass) transition temperature $T_{SG}(D)$ should be a regular function of the anisotropy $D$. If the random anisotropy is invariant under $D\leftrightarrow -D$, the spin-glass transition temperature $T=T_g$ would depend on the anisotropy $D$ as $T_g(D)\sim T_{{\rm CG}}(0)+cD^2 + \cdots $ ($c$ is a numerical constant), while, if the random anisotropy is of the type without such a  $D\leftrightarrow -D$ invariance,  the spin-glass transition temperature would depend on the anisotropy $D$ as $T_{SG}(D)\sim T_{{\rm CG}}(0)+cD + \cdots $. In case of canonical spin glasses, the main source of the random anisotropy is the DM interaction, which seems to possess the  $D\leftrightarrow -D$ symmetry.

 In any case, the spin-glass transition of the weakly anisotropic Heisenberg-like system is dictated by the chiral-glass fixed point of the isotropic system. Hence,  the spin-glass critical exponents of Heisenberg-like spin-glass magnets are given by the set of chiral-glass exponents of the isotropic system $\beta \simeq 1$, $\gamma \simeq 2$, $\delta \simeq 3$ and $\eta \simeq 0.6$, which, as we have already seen, differ significantly from the 3D Ising spin-glass values. Furthermore, even for the weakly anisotropic spin-glass magnets, the Heisenberg-to-Ising crossover is not expected in its critical behavior.

 As emphasized in the subsection C above, the spin-chirality decoupling of the fully isotropic system is a long-scale phenomenon expected to occur beyond a certain crossover length scale $l_\times$ (MC yields $l_\times \simeq 10-20$ lattice spacings). Namely, the isotropic system exhibits a crossover from the short-scale spin-chirality coupling behavior at $l < l_\times$, roughly described by $\chi \sim S^3$, to the long-scale spin-chirality decoupling behavior at $l > l_\times$. Likewise, the spin-chirality recoupling phenomenon of the weakly anisotropic system is also a long-scale phenomena  expected to occur beyond $l_\times$. The  length crossover of the weakly anisotropic system is from the short-scale spin-chirality coupling behavior at $l < l_\times$, roughly described by $\chi \sim S^3$, to the long-scale spin-chirality recoupling behavior at $l > l_\times$, roughly described by $\chi \sim S$. 


\subsection{Effects of magnetic fields}

In this subsection, on the basis of the spin-chirality decoupling-recoupling scenario developed in the previous subsections, we study the effect of magnetic fields on the spin-glass ordering, particularly with interest in the phase diagram of Heisenberg-like spin-glass magnets in the temperature - magnetic field plane. We begin with the fully isotropic Heisenberg case, and then proceed to the more realistic case with weak random magnetic anisotropy.

\subsubsection{Isotropic case}

 In the fully isotropic case, the symmetry of the Hamiltonian reduces from $Z_2\times SO(3)$ in zero field to $Z_2\times SO(2)$ under fields, where $Z_2$ refers to the chiral degeneracy associated with a spin-reflection operation (solely in spin space, not in real space) with respect to an arbitrary plane in spin space including the magnetic-field axis, while $SO(2)$ refers to the continuous degeneracy associated with a spin-rotation operation (in spin space, not in real space) around the magnetic-field axis in spin space. 

 Since the $Z_2$ chiral symmetry characterized by the sign of the scalar chirality remains under magnetic fields, the chiral-glass transition is expected to persist under magnetic fields. Of course, applied fields change the symmetry, but lower only the continuous part from $SO(3)$  to $SO(2)$. Since the continuous part has already been decoupled from the discrete $Z_2$ part, a natural expectation here would be that the chiral-glass transition persists even under fields essentially of the same type as the zero-field one. In particular, the chiral-glass transition line under fields should be a regular function of the filed intensity $H$. Since there is a trivial $H\leftrightarrow -H$ symmetry, the chiral-glass transition temperature under fields should behave for weak fields as $T_{CG}(H)\approx T_{CG}(0)- cH^2 \cdots $ ($c$ is a constant). In fact, this yields a transition line resembling the so-called GT line of the mean-field model, $|T_{CG}(0)-T_{CG}(H)|\propto H^{1/2}$, although the origin of the exponent $1/2$ is entirely different: Here, 1/2 is just of regular origin, whereas the exponent $1/2$ in the mean-field model cannot be regarded as of regular origin. The expected phase diagram expected from the chirality scenario is sketched in Fig.13(a). A similar phase diagram was obtained for the 3D isotropic Heisenberg EA model by MC simulations \cite{KawaIma}.

\begin{figure}[t]
\includegraphics[scale =0.35]{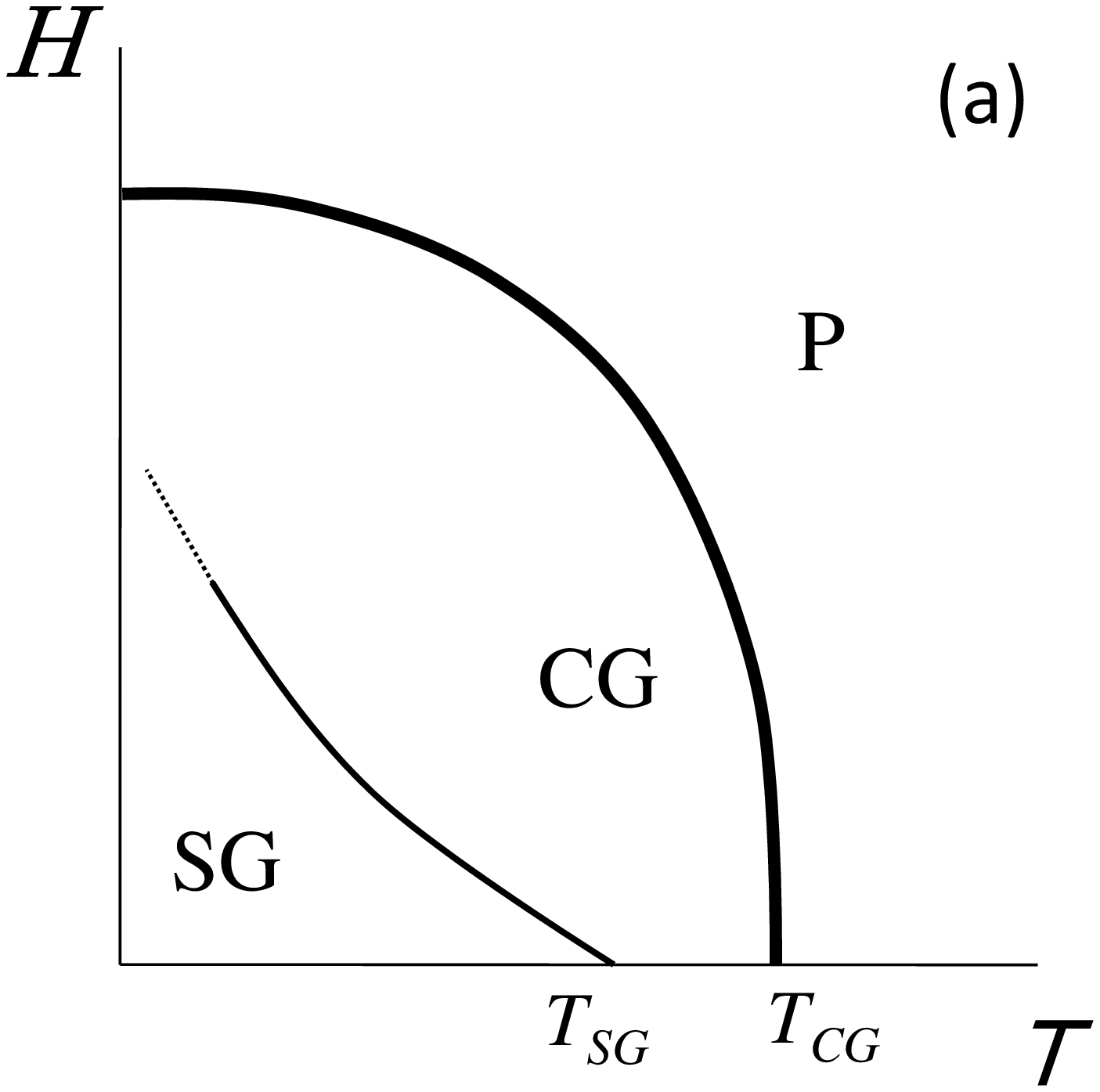}
\includegraphics[scale =0.35]{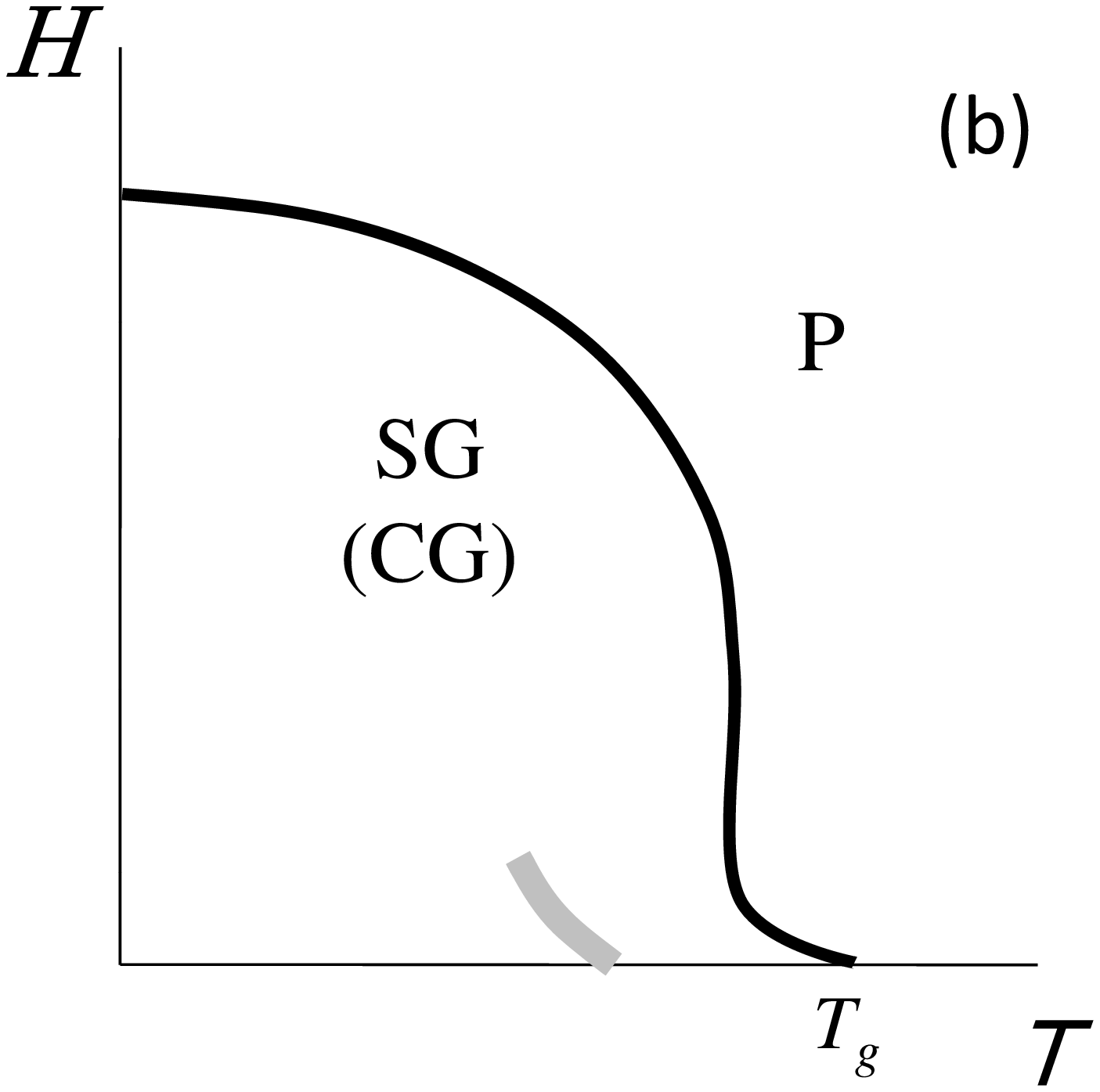}
\caption{
 Typical phase diagram of Heisenberg-like spin glasses in the magnetic field ($H$) - temperature ($T$) plane, expected from the chirality scenario. The case (a) corresponds to the fully isotropic system, and the case (b) corresponds to the more realistic weakly anisotropic system. The solid line represents a true transition line, while the thin shaded line represents a crossover line. `P', `SG' and `CG' represent the paramagnetic, spin-glass and chiral-glass phases, respectively.
}
\end{figure}

 The ordering associated with the continuous part should also occur under finite fields. The in-field transition line associated with the continuous $SO(2)$ symmetry breaking should be a continuation of the $SO(3)$ breaking spin-glass transition in zero field. Since the broken symmetry is different in zero and finite fields, {\it i.e.\/}, $SO(3)$ versus $SO(2)$, the spin-glass transition line at low fields should exhibit a singular form, $H\propto |T_{SG}(0)-T_{SG}(H)|^{\phi/2}$. The exponent $\phi$ is not yet precisely determined, but may roughly be estimated as $\phi=\beta_{SG}+\gamma_{SG}\approx 4$. This $SO(2)$ breaking transition line is also sketched in Fig.13(a).

\subsubsection{Anisotropic case}

 Now, we consider the spin-glass ordering under magnetic fields for more realistic case of the 3D Heisenberg spin glass with random magnetic anisotropy. In the presence of both random anisotropy and magnetic fields, all global symmetries of the Hamiltonian are lost at the Hamiltonian level. The chiral-glass transition associated with global $Z_2$ chiral symmetry, which was possible in the fully isotropic case,  is then no longer possible. In standard ferromagnets, such an absence of any global symmetry results in the absence of any second-order transition.  In the present case, however,  since the spin-glass (simultaneously chiral-glass) ordering in zero field accompanies RSB in addition to the breaking of the global $Z_2$ symmetry, the spin-glass (simultaneously chiral-glass) transition associated with RSB is still possible under magnetic fields. Hence, we expect that a true thermodynamic spin-glass (chiral-glass) transition might persist as an RSB transition even under magnetic fields. 

 At sufficiently high magnetic fields where the field energy overwhelms the random anisotropy energy, the behavior of the transition line would be described by that of the fully isotropic system discussed above. Let us normalize the temperature $T$, the field intensity $H$, and the magnetic anisotropy $D$ by $k_BT_g(H=0)$ which is a quantity of $O(J)$ ($J$ is a typical exchange energy) as $t=|(T_g(0)-T_g(H))/T_g(0)|$, $h=H/(k_BT_g(H=0))$ and $d=D/(k_BT_g(H=0))$. The transition line at higher fields would behave as
\begin{equation}
h \approx ct^{1/2}
\end{equation}
where $c$ is a constant. 

 At lower fields where the random anisotropy  overwhelms the magnetic-field, the transition line would behave very differently. As discussed above, the broken symmetry under finite fields (RSB only) is different from that in zero field (RSB and $Z_2$ chiral symmetry), which means that the spin-glass (simultaneously chiral-glass) transitions in each case should be  described by different fixed points. In such a situation, one expects that an applied field induces a true {\it crossover\/} phenomenon between the two distinct fixed points. Since the ordering field of the spin-glass transition is $H^2$, a singular part of the free energy $f_s$ of the weakly anisotropic spin glass is expected to have the following scaling form,
\begin{equation}
f_s \approx t^{2-\alpha} f(\frac{h^2}{t^{\beta + \gamma}}) = t^{2-\alpha_{CG}} f(\frac{h^2}{t^{\beta_{CG} + \gamma_{CG}}}),
\end{equation}
where $f$ is a scaling function, $\beta_{CG}\sim 1$ and $\gamma_{CG}\sim 2$ are the chiral-glass exponents describing the zero-field chiral-glass transition of  the isotropic system, which, according to our argument, should give the spin-glass exponents of the weakly anisotropic Heisenberg-like spin glass.

 From this scaling relation, the transition line in sufficiently low fields is expected to behave as
\begin{equation}
h \approx ct^{({\beta_{CG}}+{\gamma_{CG}})/2} \approx ct^{3/2}. 
\end{equation}
An interesting observation here is that the exponent appearing in eq.(7) is $(\beta_{CG}+\gamma_{CG})/2$, which happens to be very close to the corresponding value of the AT line of the mean-field model, 3/2. Of course, since we do not have any reason to expect that the exponent is exactly 3/2, the coincidence observed here is largely accidental. However, the expected difference is fairly small numerically, which makes the low-field transition line having an appearance of the AT-line of the mean-field theory. 

 In this way, as sketched in Figs.2 and 13, the behavior of the transition line and the magnetic phase diagram expected from the chirality scenario turn out to be quite similar to those of the mean-field theory in the entire field regime. In particular, we expect the ``GT line'' at higher fields and the ``AT line'' at lower fields, though their origin is entirely different from the mean-field GT and AT lines. 

 In the fully isotropic case, as argued in the previous subsection, the spin-chirality decoupling is expected to persist even under fields so that the $SO(2)$ symmetry-breaking transition line appears at a temperature lower than the $Z_2$ symmetry-breaking chiral-glass transition line, as shown in Fig.13(a). In the present weakly anisotropic case, due to the spin-chirality recoupling, this $SO(2)$ symmetry-breaking transition line should be smeared out: It can not remain as a true transition line. Its remnant, however, may still be observable as a sort of crossover line {\it if the anisotropy is sufficiently weak\/}. If the anisotropy is not weak enough, this crossover line will be rapidly smeared away, or will be driven toward the higher-temperature chiral-glass (simultaneously spin-glass) transition line due to the effect of the spin-chirality recoupling. 
%
%
This  smeared transition line (or crossover line) is also sketched in Fig.13(b). Interestingly, this smeared transition line somewhat resembles the so-called ``high-field AT line'' sometimes observed experimentally in canonical spin glasses \cite{Kennig}.  Thus, our present scenario gives a new perspective in explaining why the mean-field theory appears to be so good in describing the experimental phase diagram.

 In-field ordering properties of the weakly anisotropic Heisenberg spin glass were also studied by Imagawa and the author by means of MC simulations, which successfully demonstrates some of the feature of the phase diagram mentioned above \cite{ImaKawa}. Another interesting observation from MC is that the spin-glass ordered state turns out to be quite robust against applied magnetic fields \cite{KawaIma,ImaKawa}. It appears to be stable up to fields as large as 25$k_B T_{SG}(H=0)$. This might be understandable if one notices that the coupling between the chirality and magnetic fields might be rather weak, since magnetic fields couple directly to the spin via the Zeeman term, only indirectly to the chirality.

\subsection{Replica-symmetry breaking}

 Numerical simulations on the 3D isotropic Heisenberg spin glass suggest that the chiral-glass ordered state is non-self-averaging  \cite{HukuKawa00,HukuKawa05,VietKawamura}. It was first suggested by Hukushima and the author that the chiral-glass ordered state might accompany a one-step-like RSB \cite{HukuKawa00,HukuKawa05,VietKawamura}, in sharp contrast to the cases of the 3D Ising spin glass or of the mean-field SK model. Chirality scenario then expects that the spin-glass ordered state exhibits essentially the same type of one-step-like RSB also in the spin sector due to the spin-chirality recoupling. Such a one-step feature should be reflected most notably in a sharp central peak in the associated overlap-distribution function $P(q)$. 

 As has been explained in \S IIC, recent extensive studies have revealed that the overlap distribution function $P(q)$ is related to the so-called off-equilibrium fluctuation dissipation ratio $X(C)$, the ratio between the off-equilibrium response (susceptibility) $\chi$ and the off-equilibrium autocorrelation function $C$. Typical behaviors of the so-called $\chi -C$ plots expected in the spin-glass ordered state have been given in Fig.4, together with experimental data of the Heisenberg-like spin-glass magnet in Fig.5. If there occurs a one-step-like RSB, the $\chi-C$ curve should approach the $C=0$ axis with a finite (nonzero) tangent, while, if the spin-glass ordered state accompanies no RSB or the full RSB, the $\chi-C$ curve should approach the $C=0$ axis with a vanishing tangent. 

 The  $\chi-C$ curve was also calculated numerically for the weakly anisotropic 3D Heisenberg spin glass with the  $\pm J$ couplings \cite{Kawamura03}, and the result is reproduced in Fig.14. The system size is $32^3$, and the data are given for several choices of the waiting time $t_w$ at the bath temperature well below the spin-glass (simultaneously chiral-glass) transition temperature $T_g$. The data suggest that there occurs two-stage dynamical process: In the quasi-equilibrium regime $t<<t_w$, the $\chi$-$C$ plot exhibits a linear behavior satisfying the fluctuation-dissipation relation, while, in the off-equilibrium regime at $t>>t_w$, the  $\chi$-$C$ plot exhibits another linear behavior characterized by the ``effective temperature'' $T_{eff}$ which comes around $T_{eff}\simeq 2T_g$ irrespective of the bath temperature.  The latter feature is consistent with the one-step-like  behavior of Fig.4(c) or (d) as expected from the chirality scenario.

\section{Relation to experiments and discussion}

 In this section, we wish to discuss the present  experimental situation of the spin-glass ordering from the standpoint of the chirality theory. This issue is discussed in detail by I.A. Campbell and D. Petit in a companion paper of this volume. Hence, I will discuss the point very briefly here.

 Let us first examine the three points which I raised in \S 2 and discussed in \S 3 from the viewpoint of the chirality theory.

\subsection{The problem of the critical properties}

 According to the chirality scenario, the spin-glass critical exponents of canonical spin glasses, or more generally the weakly anisotropic Heisenberg-like spin glasses, should be given by those of the chiral-glass exponents of the fully isotropic system, which are respectively given by $\alpha\simeq -2.5$, $\beta\simeq 1.2$, $\gamma\simeq 2.1$, $\nu\simeq 1.5$ and $\eta\simeq 0.6$. Note that these exponents are totally different from the corresponding Ising spin-glass values \cite{Fert,Simpson,Coutenary86,Bouchiat,Levy,Coles,Taniguchi88,Campbell09}. These predictions are compared quite favorably with the available experimental data, which provides the strongest support to the chirality scenario.

\subsection{The problem of the magnetic phase diagram of spin glasses}

 According to the chirality scenario, a true thermodynamics spin-glass transition should persist as an RSB transition even under finite magnetic fields. At low magnetic fields, the spin-glass transition line behaves as the $H/(k_BT_g(0))\approx c|(T_g(0)-T_g(H))/T_g(0)|^{(\beta_{CG}+\gamma_{CG})/2} \approx c|(T_g(0)-T_g(H))/T_g(0)|^{3/2}$, while at higher magnetic fields it behaves as  $H\approx c|(T_g(H)-T_g(0))/T_g(0)|^{1/2}$. These transition lines are originated from the chiral-glass fixed point of the fully isotropic system. Although these behaviors of the transition line in fields have some resemblance to those of the mean-field model, {\it i.e.\/} the so-called  AT and GT lines, the physical origin is very different, and such resemblance is largely accidental. 
%
%
In addition, the chirality scenario expects, for systems with sufficiently weak random magnetic anisotropy, a smeared transition line or a crossover line (not a true transition line) at a lower temperature, which behaves as $H/(k_BT_g(0))\approx c|(T_g(0)-T_g(H))/T_g(0)|^{\phi/2}$ ($\phi$ is roughly estimated as $\phi\sim 4$). This crossover line is originated from the spin-glass fixed point of the fully isotropic system, a different one from the chiral-glass fixed point. It might be related to the high-field AT-like line sometimes observed experimentally below the GT-like transition line \cite{Kennig}. Thus, the chirality scenario provides a new consistent explanation of the experimentally observed phase diagram of Heisenberg-like spin glasses from an entirely different perspective from the mean-field theory \cite{Campbell09,Campbell99,Campbell02,Chamberlin,Courtenary84,Kennig}.

\subsection{The problem of RSB in the spin-glass ordered state}

 According to the chirality scenario, the spin-glass ordered state of the weakly anisotropic Heisenberg spin glass is non-self-averaging, and exhibits a one-step-like RSB. Such a one-step feature might be observable via a characteristic $\chi -C$ relation in its off-equilibrium dynamics. As already shown in Fig.5, the experimental data available so far seems consistent with such a one-step-like behavior expected from the chirality scenario \cite{HO}. It might be interesting to perform further experimental studies  to determine the asymptotic behavior of the  $\chi-C$ curve in the vicinity of $C=0$, and to extend to other types of spin-glass materials, particularly to metallic canonical spin glasses.

\begin{figure}[t]
\includegraphics[scale =0.8]{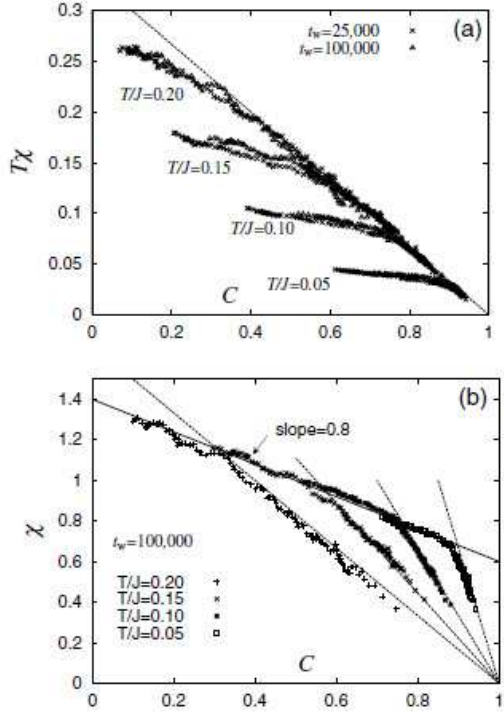}
\caption{
The susceptibility, $T\chi$ (a) or $\chi$ (b), versus correlation $C$ plot of the weakly anisotropic Heisenberg EA model with $D/J=0.01$ at several temperatures below $T_g\simeq 0.21J$, where $J$ and $D$ represent the magnitudes of the isotropic exchange interaction of the $\pm J$-type and of the random magnetic anisotropy distributed uniformly between [$-D$, $D$]. The applied field intensity is $H/J=0.01$. The lattice size is $L=32$. The broken lines in (b) represent the FDT lines. The straight line in (b) is the straight-line fit to the data in the aging regime, its slope being equal to 0.8 which can be  translated to the relation that the effective temperature governing the aging dynamics is $T_{eff}\simeq 2T$ irrespective of the bath temperature $T$.  [From H. Kawamura, Phys. Rev. Letters {\textbf 90} (2003) 237201.]
}
\end{figure}

\subsection{Direct measurements of the chirality}

 The most stringent experimental test of the chirality scenario would be to directly measure the chirality, particularly, the chiral susceptibility $X_\chi$ and the nonlinear chiral susceptibility $X_{\chi}^{nl}$. This has long remained to be a difficult task, since the chirality is a higher-order quantity in spins, making its experimental detection rather difficult. Recently, however, it has been recognized that the scalar chirality might be measurable by using the anomalous Hall effect as a probe. In fact, G. Tatara and the present author analyzed the chirality contribution to the anomalous Hall effect of metallic spin glasses based on the perturbation analysis \cite{TataraKawamura,KawamuraHall}. The anomalous Hall coefficient $R_s$, which is the ratio of the Hall resistivity $\rho_H$ and the magnetization $M$, is then given by
\begin{eqnarray}
R_s &=& \rho_H / M \nonumber\\
    &=& - \left( A\rho + B\rho^2 \right) - CD \left( X_\chi + X_{\chi}^{nl}(DM)^2 + \cdots \right). 
\end{eqnarray}
It consists of two kinds of terms. The first part is the skew and the side-jump contributions to the anomalous Hall effect which are proportional to the resistivity $\rho$ or its squared $\rho^2$. Since the longitudinal resistivity $\rho$ of spin glasses does not show any anomaly at $T_g$, this first part can be regarded as a regular background. The second part is the chirality contribution, which is proportional to the chiral susceptibility $X_\chi$. It even contains the information  of the nonlinear chiral susceptibility  $X_{\chi}^{nl}$ as a higher-order contribution. 

 Inspired by this theoretical suggestion, several experimental groups tried to measure the chirality contribution to the anomalous Hall effect in metallic spin glasses \cite{Sato,Taniguchi04,Campbell04,Campbell06,Taniguchi07,Yamanaka07}. These measurements observed a sharp cusp-like anomaly at $T=T_g$ in the temperature dependence of $R_s$, followed by the deviation between the field-cooled and the zero-field-cooled data below $T_g$ \cite{Taniguchi04,Campbell04}.  Taniguchi {\it et al\/} recently observed a singular behavior of the nonlinear chiral susceptibility at $T=T_g$ characterized by the exponent $\delta_{{\rm CG}}\simeq 3$, which is rather close to the corresponding chiral-glass exponent determined numerically \cite{Taniguchi07}. As an example, we reproduce in Fig.15 the experimental data of the Hall coefficient taken from Ref.\cite{Taniguchi04}.

\begin{figure}[t]
\includegraphics[scale =1.0]{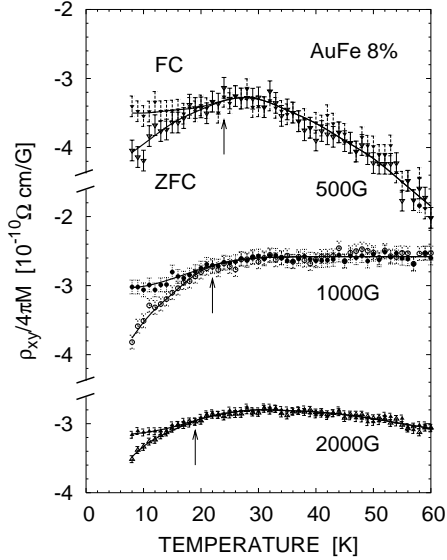}
\caption{
The temperature dependence of the Hall coefficient $\rho/M$ of canonical spin glass AuFe 8\% in applied fields, where $\rho$ is the Hall resistivity and $M$ is the magnetization. Arrows in the figure indicate the freezing temperature $T_f(H)$. [From T. Taniguchi {\it et al\/}, Phys. Rev. Letters {\textbf 93} (2004) 246605.]
}
\end{figure}

 We stress that, if the order parameter of the spin-glass transition were not the  chirality but were the spin itself as in the case of the mean-field Heisenberg SK model,  the chiral susceptibilities would not exhibit a strong singularity. For example, the nonlinear chiral susceptibility of the Heisenberg SK model does not diverge at $T_g$ \cite{ImaKawa03}. This is simply due to the fact that the chirality is a composite operator, being of higher-order in the spin. In the absence of the spin-chirality decoupling, we expect $\nu_{CG}=\nu_{SG}$, and from a simple power-counting argument, we have $\beta_{CG}\simeq 3\beta_{SG}$ as a first-order approximation, and from a scaling relation, $\gamma_{CG}\simeq \gamma_{SG}-4\beta_{SG}$, {\it etc\/}. If one substitutes here the experimental spin-glass exponents for canonical spin glasses, $\beta_{{\rm SG}}\simeq 1$ and $\gamma_{{\rm SG}}\simeq 2$, one gets the chiral-glass susceptibility exponent $\gamma_{{\rm CG}}\simeq -2$. Hence, if there is no spin-chirality decoupling and the spin remains to be a primary order parameter of the transition, one would not expect any discernible anomaly in the nonlinear chiral susceptibility, quite contrary to the recent experimental observation \cite{Sato,Taniguchi04,Campbell04,Campbell06,Taniguchi07,Yamanaka07} !

 Hence, a distinct anomaly observed in the Hall coefficient measurement of canonical spin glasses indicate that the chiral susceptibility or the nonlinear chiral susceptibility in metallic spin glass indeed exhibits a prominent anomaly at the spin-glass transition, thereby providing a strong experimental support to the chirality scenario of the spin-glass ordering.

\section{Other related systems}

In this section, I will take up several related systems other than the 3D Heisenberg spin glass which might exhibit the spin-chirality decoupling phenomena or its correspondence. These include, (A) vector spin glasses in one and two dimensions, (B) the one-dimensional Heisenberg spin glass with the long-range power-law interaction, (C) regularly frustrated {\it XY\/} models in one and two dimensions, and (D) granular cuprate superconductors and the {\it XY\/} spin glass in three dimensions. I wonder if these examples might convey the feeling to the reader that the chiral order is really a general and far-reaching concept, not solely limited to the Heisenberg-like spin glass, possibly opening a new horizon in condensed matter physics.

\subsection{The ordering of vector spin glasses in two dimensions}

 The ordering of two-dimensional (2D) vector spin glasses not only has its own interest but also gives us a hint to understand the behavior of the 3D system.  Since an experimental realization of the 2D spin-glass material has been scarce so far, many of the studies on the 2D spin glass have been numerical ones. Both the three-component Heisenberg spin-glass \cite{KawaYone} and the two-component {\it XY\/} spin-glass models \cite{KawaTane,Batrouni,KawaTane2,RayMoore,KawaXY95,Bokil,Wengel,Grempel,Kosterlitz,Granato2D,Weigel} have been studied. In case of the {\it XY\/} spin glass, The spin is a two-component vector $S$=($S_x,S_y$)=($\cos\theta,\sin\theta$), and the chirality is a vector chirality $\kappa_i=S_i^xS_{i+\delta}^y - S_i^yS_{i+\delta}^x = \sin (\theta_i - \theta_{i+\delta})$. 

 In both {\it XY\/} and Heisenberg spin glasses in 2D, there now appears a consensus that the chirality and the spin order only at $T=0$. Nevertheless, it was suggested first by the present author and Tanemura that the spin and the chirality of the 2D vector spin glass might be decoupled in the sense that there exist two distinct diverging length scales in the $T=0$ transition, one associated with the chirality $\xi_{CG}$ and the other associated with the spin $\xi_{SG}$, with $\xi_{CG} >> \xi_{SG}$ \cite{KawaTane,KawaTane2,KawaYone}. More precisely, the spin correlation-length exponent $\nu_{SG}$ and the chiral correlation-length exponent $\nu_{CG}$ characterizing the $T=0$ transition are mutually different, and one has $\nu_{CG} > \nu_{SG}$.  

 Indeed, numerical simulations support this suggestion. Most calculations yield the chiral-glass exponent around $\nu_{CG}\sim 2$ and the spin-glass exponent around $\nu_{CG}\sim 1$, hence $\nu_{CG} > \nu_{SG}$ \cite{KawaTane,Batrouni,KawaTane2,RayMoore,KawaXY95,Bokil,Wengel,Grempel,Granato2D,KawaYone,Weigel}. In particular, recent `almost exact' calculation by Weigel and Gingras on the 2D {\it XY\/} spin glass for finite lattices up to $28^2$ spins has shown that there are indeed two different sets of stiffness exponents, one for the spin excitation and the other for the chiral excitations, demonstrating that the spin and the chirality are really decoupled in the 2D {\it XY\/} spin-glass model \cite{Weigel}.

 Some theoretical works have also been made for the {\it XY\/} spin-glass model on a 1D ladder lattice \cite{Ney,Uda}. Again, `almost exact' analysis is possible. It is shown that the $Z_2$ chiral and $SO(2)$ continuous (or spin-wave) degrees freedom are decoupled in the sense that the full spin correlation function can be written as a product of the two parts, one is the chirality-chirality correlation function and the other is the spin-wave correlation function. However, a twist occurs here in that the spin (spin-wave) correlation-length exponent $\nu_{SG}=1$ happens to be greater than the chirality correlation-length exponent $\nu_{CG}=0.5263...$, in contrast to the 2D and 3D cases. Because of this inequality, the spin-spin correlation function is eventually characterized by the chiral correlation-length exponent $\nu_{CG}$ instead of  $\nu_{SG}$. In this way, the spin-chirality decoupling phenomena has been ``masked'' in this 1D model \cite{Uda}.

 We note that possible spin-chirality decoupling  of the Heisenberg EA model was also examined numerically in dimensions greater than $D=3$ \cite{ImaKawa03}. It was suggested that the decoupling might still occur at least in four dimensions ($D=4$), although the decoupling eventually goes away in high enough dimensions.

\subsection{The ordering of the one-dimensional Heisenberg spin glass with the long-range power-law interaction}

 Even in one dimension,  a finite-temperature transition becomes possible when the interaction becomes sufficiently long-ranged. In this connection, the ordering properties of the one-dimensional spin-glass model with the long-range power-law interaction proportional to $1/r^{\sigma}$ ($r$ is the spin distance) is of interest \cite{Katzgraber04,Matsuda,Katzgraber09}. In the limit of sufficiently large $\sigma\rightarrow \infty$, the model reduces to the standard 1D model with short-range interaction, while, in the opposite limit of $\sigma\rightarrow 0$, the model should reduce to an infinite-range mean-field model which corresponds to $D=\infty$. (In fact, $\sigma =1/2$ is another boarder-line value below which one needs to introduce some appropriate normalization procedure to keep the energy extensive.) One may then expect that varying the power-law exponent $\sigma$ might roughly correspond to varying the dimensionality $D$ in the corresponding to {\it short-range\/} model. Obviously, the system exhibits only a $T=0$ transition in the $\sigma\rightarrow \infty$ limit, while it exhibits a finite-temperature transition of the mean-field type in the $\sigma \rightarrow 0$ limit. In fact, a recent numerical study of the 1D Ising spin-glass model with the long-range power-law interaction gives some support to such correspondence between $\sigma$ and $D$ \cite{Katzgraber04}.

 An interesting question arises here concerning whether the spin-chirality decoupling is ever realized in the corresponding Heisenberg spin-glass model for certain range of $\sigma$. Since it has been established that the mean-field Heisenberg SK model does not show the spin-chirality decoupling, the spin-chirality decoupling accompanied with a finite-temperature transition should be realized, if any, only in a certain intermediate range of $\sigma$. Renormalization-group calculation, which did not take account of the possible effect of the spin-chirality decoupling, suggested that $\sigma=1$ was a borderline value separating a zero-temperature transition regime and a finite-temperature transition regime (corresponding to the lower critical dimension) \cite{BMY,Kotliar1D,Chang}. Since recent numerical consensus is that there exists a finite-temperature spin-glass transition in $D=3$, one may deduce that the value of $\sigma$ corresponding to the physical dimension of the short-range model lies slightly below $\sigma=1$. Of course, one should keep in mind that such a $\sigma$-$D$ correspondence is only empirical at the present stage without firm theoretical basis. Moreover, the RG observation that $\sigma=1$ lies just at a border is based an argument neglecting the possibility of the spin-chirality decoupling. Once this possibility has been taken into account, the true borderline value of $\sigma$ might change.

 The spin and the chirality orderings of the 1D Heisenberg spin glass with the long-range power-law interaction has recently been studied in a wide range of $\sigma$ by extensive MC simulations by Viet and the author (unpublished work). An advantage in numerically studying such a 1D model might be twofold: Firstly, relatively large system can be simulated in 1D (up to $L=2048$).  Secondly, one can continuously change  and even fine-tune the parameter $\sigma$, while it is totally impossible to continuously change the dimensionality $D$. Recent MC then indicates that the expected spin-chirality decoupling is realized in the range $0.8\lsim \sigma \lsim 1.1$. Numerical evidence of the spin-chirality decoupling in this model is particularly clear at, say, $\sigma=0.95$, being much clearer than the one observed in the short-range model in 3D. Such a strong numerical evidence has become possible because one can continuously vary and optimize the parameter $\sigma$. Hence, the result of the 1D long-range model lends indirect support to the view that the spin-chirality decoupling might well occur also in the 3D short-range model. Some of the features of the 1D long-range model, however, appear to be not completely the same as those of the 3D short-range model. For example, a one-step RSB feature of the chiral-glass ordered state observed in the 3D short-range model is very weak or might be absent here even in the regime of $\sigma$ exhibiting the spin-chirality decoupling.

\subsection{The spin and the chiral orderings of regularly frustrated vector spin models}

 Next, we review briefly the spin and the chirality orderings of regularly frustrated {\it XY\/} antiferromagnets. In case of regularly frustrated vector spins, the ordered state spin structure becomes noncollinear but often coplanar. It means there is a nontrivial vector chirality, but no nontrivial scalar-chirality. Reflecting this situation, {\it XY\/} spin models are often considered in {\it regularly frustrated\/} models since the vector chirality leads to a $Z_2$ discrete chiral degeneracy in the {\it XY\/} case.

 First example is the classical {\it XY\/} (plane rotator) model on the one-dimensional (1D) triangular-ladder lattice. Since the model is one-dimensional, there cannot be any finite-temperature transition. An advantage here is that the model is exactly solvable \cite{Horiguchi}. While both the spin and the chirality order only at $T=0$, the associated correlation-length exponents characterizing the $T=0$ transition are mutually different.  Indeed, the spin correlation-length exponent is equal to unity $\nu_s = 1$, while the chiral correlation-length exponent is equal to $\nu_\kappa = \infty$, meaning that the chiral correlation length diverges exponentially toward $T=0$ \cite{Horiguchi}, $\xi({\rm chirality}) >> \xi({\rm spin})$. Hence, in this particular 1D model, the spin-chirality decoupling is rigorously shown to occur in the sense that there exist two distinct diverging length scales in the transition though the transition occurs only at $T=0$.

 Second example is the classical {\it XY\/} (plane rotator) antiferromagnet on the 2D triangular lattice or on the 2D fully frustrated square lattice (the so-called odd lattice). Although there had been  some controversy concerning how the chiral $Z_2$ and the spin-rotation $SO(2)$ degrees of freedom order in these systems, consensus now appears that separate spin and chirality transitions occur successively at finite temperatures \cite{MiyashitaShiba,Xu,Capriotti,Lee,Loison,Ozeki,Hasenbusch05}. With decreasing the temperature, the chirality orders first at a higher temperature into the long-range ordered state, while the spin orders at a lower temperature into the quasi-long-range ordered state. The mutual distance between the chirality transition and the spin transition is rather close, only about 1\% difference in the temperature. 

 Another example may be the frustrated classical {\it XY\/} (plane rotator) model on the 2D square lattice with a fractional external flux threading the system whose strength is irrational to the lattice periodicity. It has been found that this model exhibits a $T=0$ transition, which are characterized by two distinct diverging correlation lengths: One is associated with the externally generated vortex corresponding to the chirality, and the other is associated with the spin itself \cite{Granato}. So, the spin-chirality decoupling occurs at the $T=0$ transition of this model. 

 Even more interesting situation seems to occur when the coupling of this model is modified to be spatially anisotropic, {\it i.e.\/}, the coupling along the $x$-direction $J_x$ is taken to be different from that in the $y$-direction $J_y$. Very recent simulation by Yoshino {\it et al\/} has revealed that the chirality (externally introduced vortex) exhibits a finite-temperature transition without the conventional spin order, exhibiting a remarkably prominent spin-chirality decoupling phenomenon \cite{Yoshino09}.

\subsection{The ordering of granular cuprate superconductors and the {\it XY\/} spin glass}

 Another vector spin-glass model possessing the nontrivial $Z_2$ chiral degrees of freedom is the {\it XY\/} spin glass. As in case of the Heisenberg spin glass, the possibility of the spin-chirality decoupling has also been suggested, {\it i.e.\/}, successive chiral-glass and spin-glass transitions occurring at $T=T_{CG}$ and at $T=T_{SG}$ with $T_{CG} > T_{SG}$ \cite{KawaTane3DXYSG}. Some numerical support of the decoupling was reported \cite{Kawa3DXYSG,KawaLi3DXYSG}, but some other groups claimed a simultaneous occurrence of the spin and the chiral transitions \cite{Granato3D,Nakamura04,PixleyYoung08}. Hence, the subject still remains somewhat controversial. 

 Experimental realization of the {\it XY\/} spin glass might be found in certain spin-glass magnets with an easy-plane-type uniaxial anisotropy, {\it e.g.\/}, Rb$_2$Mn$_{1-x}$Cr$_x$Cl$_4$ \cite{Katsumata}, CdMn \cite{Murayama} and Eu$_{0.5}$Sr$_{1.5}$MnO$_4$ \cite{Mathieu}. As was first pointed out in Refs.\cite{Kawa-orbital,KawaLi}, other interesting experimental realization might be certain granular cuprate superconductors consisting of random Josephson network of sub-micron-size superconducting grains \cite{Matsuura}. Since cuprate superconductors are $d$-wave superconductors, Josephson junction between two cuprate crystallites with random spatial orientations can be either ``$\pi$-junction'' or ``0-junction'' according as the phase of the superconducting order parameter is shifted by $\pi$ or 0 across the junction.

\begin{figure}[t]
\includegraphics[scale =0.5]{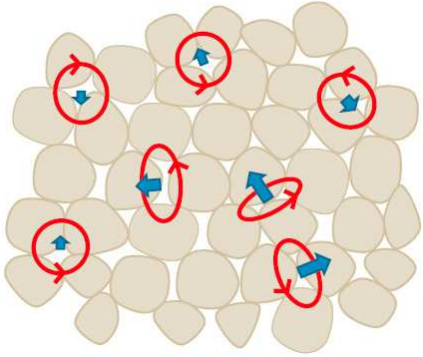}
\caption{
(Color online) Illustration of the chiral-glass ordered state in granular cuprate superconductors. Circle represents a loop-supercurrent spontaneously flowing either in clockwise or counterclockwise direction, while arrow represents the direction of magnetic flux induced by the loop-supercurrent. 
}
\end{figure}

 As is known, macroscopic properties of the Josephson network is well described by the {\it XY\/} model phase Hamiltonian, $H = -\sum _{<ij>} J_{ij} \cos(\theta_i - \theta_j)$, where the random interaction $J_{ij}$ is either `ferromagnetic' representing the 0-junction, or `antiferromagnetic' representing the $\pi$-junction \cite{Kawa-orbital}. Existence of both 0 and $\pi$ junctions inevitably gives rise to the frustration effect. Hence, a close analogy to the problem of the 3D {\it XY\/} spin glass arises.  In reality, superconductors are charged and there appears an additional coupling to fluctuating magnetic fields (the screening effect) \cite{KawaLi}. 
 
 The chirality and the spin in magnets have interesting counterparts in Josephson networks \cite{Kawa-orbital,KawaLi}. Namely, the spin in magnets corresponds to the phase of the superconducting order parameter (Cooper-pair wavefunction) at each grain, while the chirality in magnets corresponds to the circulation of superconducting current-loop flowing in Josephson network. Nonzero chirality in magnets corresponds to nonzero superconducting loop-current in Josephson network spontaneously flowing either in clockwise or in counter-clockwise direction. Since the circulating current-loop generates a magnetic flux threading the loop, magnetic field in Josephson network serves as a ``chiral field''. In this sense, the role of magnetic field is dual between magnets and superconductors. 

  Such an analogy suggests the possible occurrence of the spin-chirality decoupling phenomenon and the appearance of the chiral-glass phase in the ordering process of granular cuprate superconductors \cite{KawaLi}. Indeed, numerical simulation by Li and the author performed on a random Josephson network model taking account of the screening effect supports such a conjecture \cite{KawaLi}. At a higher ``chiral-glass'' transition temperature, the chirality is frozen, {\it i.e.\/}, the circulation of superconducting current loops are frozen: See Fig.16. This can be observable experimentally as the negative divergence of the nonlinear susceptibility via the standard magnetic measurements \cite{Kawa-orbital,KawaLi,Matsuura}.  At a lower ``spin-glass'' transition temperature, the pseudo {\it XY\/} spin would be frozen, {\it i.e.\/}, the phase of the superconducting order parameter is frozen, and the system becomes a true superconductor. This can be observable experimentally as the onset of the vanishing linear resistivity  via the standard current($I$)-voltage ($V$) transport measurements. The chiral-glass state has a small but nonzero linear resistivity \cite{Kawa-transport}.

 Nice aspect about this system is twofold. For one, the internal space of the superconducting order parameter is the gauge space, not the spin space, whose isotropy (U(1) gauge symmetry) is completely respected in any situation, and there cannot be a phenomenon like the ``spin-chirality recoupling''. Hence, the chiral-glass phase in granular cuprate superconductors is quite robust against perturbations. For the other, the chirality in superconductors is just the circulating electric current which produces the magnetic moment threading the loop. So, the chiral-glass transition is experimentally measurable via the standard magnetic measurements, in contrast to the case of spin-glass magnets where the measurement of the chirality poses quite a tough technical problem. Meanwhile, the ``spin-glass'' transition, or the phase-freezing transition, should be observable via the standard transport measurement. It is then highly interesting to check experimentally whether the predicted ``spin-chirality decoupling'' phenomenon is ever realized in the ordering of granular cuprate superconductors.

 It has already been established for certain YBCO granular superconductors that an intergranular phase transition accompanied by the negative divergent-like anomaly in the nonlinear susceptibility occurs below the bulk superconducting phase transition temperature \cite{Hagiwara}. It remains to be seen whether the ``spin-chirality decoupling'' really occurs by the combination of careful magnetic and transport measurements. Such measurements are now underway \cite{deguchi}.

\section{Concluding remarks}

 I have reviewed the present status of the chirality scenario of experimental spin-glass transitions in some detail and have tried to further develop its consequences in terms of several recent numerical simulations and experiments. In this scenario, the spin-glass order of weakly anisotropic Heisenberg-like spin-glass magnets including canonical spin glasses are essentially chirality driven. An intriguing ``spin-chirality decoupling'' phenomenon expected to occur in fully isotropic Heisenberg spin glass plays a key role in the scenario. Experimental spin-glass ordering is essentially the chiral-glass ordering ``revealed'' via the weak random magnetic anisotropy.

 Similar spin-chirality decoupling phenomena could also occur in other frustrated systems including the two-dimensional triangular {\it XY\/} antiferromagnet, the Josephson-junction array in applied magnetic fields, and granular cuprate superconductors. Thus,  together with many recent advances  made in the area of multiferroics and transport properties where the chirality plays an essential role, the chirality concept might open a new horizon in modern condensed matter research.

\begin{acknowledgments}
The author is thankful to I.A. Campbell,  D.X. Viet, K. Hukushima,  D. Imagawa, G. Tatara, M. Nakamura and A. Matsuda for the collaboration and many useful discussion, and to T. Taniguchi, H. Yoshino, E. Vincent, M. Ocio, M. Picco, M. Sato and H. Takayama for discussion. This study was supported by Grant-in-Aid for Scientific Research on Priority Areas ``Novel States of Matter Induced by Frustration'' (19052006). We thank ISSP, Tokyo University and YITP, Kyoto University for providing us with the CPU time.
\end{acknowledgments}

\end{document}